\documentclass{aa}
\usepackage{rotating}
\usepackage{graphicx}
\usepackage{txfonts}
\begin{document}

\title{A multi-epoch {\it XMM-Newton} campaign on the core of the massive Cyg\,OB2 association\thanks{Based on observations collected with {\it XMM-Newton}, an ESA Science Mission with instruments and contributions directly funded by ESA Member States and the USA (NASA).}\fnmsep\thanks{Tables 2 and 4 are only available in electronic form at the CDS via anonymous ftp to cdsarc.u-strasbg.fr (130.79.128.5) or via http://cdsweb.u-strasbg.fr/cgi-bin/qcat?J/A+A/}}
\author{G.\ Rauw\inst{1}\fnmsep\thanks{Honorary Research Associate FRS-FNRS (Belgium)}}
\offprints{G.\ Rauw}
\mail{rauw@astro.ulg.ac.be}
\institute{Groupe d'Astrophysique des Hautes Energies, Institut d'Astrophysique et de G\'eophysique, Universit\'e de Li\`ege, All\'ee du 6 Ao\^ut, B\^at B5c, 4000 Li\`ege, Belgium} 
\date{Received date / Accepted date}
\abstract{Cyg\,OB2 is one of the most massive associations of O-type stars in our Galaxy. Despite the large interstellar reddening towards Cyg\,OB2, many studies, spanning a wide range of wavelengths, have been conducted to more clearly understand this association. X-ray observations provide a powerful tool to overcome the effect of interstellar absorption and study the most energetic processes associated with the stars in Cyg\,OB2.}{We analyse XMM-Newton data to investigate the X-ray and UV properties of massive O-type stars as well as low-mass pre-main sequence stars in Cyg\,OB2.}{We obtained six XMM-Newton observations of the core of Cyg\,OB2. In our analysis, we pay particular attention to the variability of the X-ray bright OB stars, especially the luminous blue variable candidate Cyg\,OB2 \#12.}{We find that X-ray variability is quite common among the stars in Cyg\,OB2. Whilst short-term variations are restricted mostly to low-mass pre-main sequence stars, one third of the OB stars display long-term variations. The X-ray flux of Cyg\,OB2 \#12 varies by 37\%, over timescales from days to years, whilst its mean $\log{\frac{ L_{\rm X}}{L_{\rm bol}}}$ amounts to $-6.10$.}{These properties suggest that Cyg\,OB2 \#12 is either an interacting-wind system or displays a magnetically confined wind. Two other X-ray bright O-type stars (MT91\,516 and CPR2002\,A11) display variations that suggest they are interacting wind binary systems.}
\keywords{Open clusters and associations: individual: Cyg\,OB2 -- Stars: early-type -- stars: pre-main sequence -- stars: individual: Cyg\,OB2 \#12 -- X-rays: stars}
\authorrunning{G. Rauw}
\titlerunning{A multi-epoch {\it XMM-Newton} campaign on Cyg\,OB2}
\maketitle
\section{Introduction}
The Cygnus\,OB2 association is one of the richest and most massive OB associations known in our Galaxy. These properties make it an interesting target for observations over a broad range of wavelengths from the radio to the $\gamma$-ray domain. Unfortunately, photometric and spectroscopic studies of Cyg\,OB2 in the UV (e.g.\ Herrero et al.\ \cite{HPCKV}) and optical waveband (e.g.\ Massey \& Thompson \cite{massey}, Torres-Dodgen et al.\ \cite{torres} and Hanson \cite{hanson}) are seriously hampered by the heavy and patchy absorption by dense molecular clouds. Nevertheless, the interest in this association has been boosted over the past decade by near-infrared studies, which have obtained a far more complete census of its stellar population, and by the discovery of two $\gamma$-ray sources, a GeV and a TeV source, in this direction. From 2MASS near-IR data, Kn\"odlseder (\cite{knoedlseder}) concluded that Cyg\,OB2 probably harbours about $2600 \pm 400$ OB stars, among which $120 \pm 20$ are O-stars, within an angular radius of about one degree. Although more recent studies suggest a slightly lower number of 90 -- 100 O-type stars (Com\'eron et al.\ \cite{Comeron}), Cyg\,OB2 is certainly quite rich in massive stars and contains stars as early as O3\,If$^*$ (Cyg\,OB2 \#7). Accordingly, it has been suggested that the stellar winds of this concentration of massive stars in Cyg\,OB2 could be the origin of the slightly extended, unidentified TeV source TeV\,J2032+4130 (Aharonian et al.\ \cite{Felix}, Albert et al.\ \cite{Albert}), although alternative scenarios have also been proposed (see Butt et al.\ \cite{Butt} and references therein). In the GeV energy domain, the unidentified {\it EGRET} $\gamma$-ray source 3EG\,J2033+4118 had also been proposed to be associated with interacting winds from massive stars in the core of the association (Romero et al.\ \cite{Romero}, Rauw \cite{HK}), but recent {\it Fermi}-LAT data have identified this object as a young energetic pulsar (PSR\,J2032+4127, Abdo et al.\ \cite{Abdo}).

Four Cyg\,OB2 members have been extensively studied in the radio domain. Three of them are non-thermal radio emitters (Cyg\,OB2 \#5, \#8a, and \#9, see e.g.\ De Becker \cite{DeBecker} and references therein) and are now known as binary or multiple systems hosting wind interaction regions (Kennedy et al.\ \cite{Kennedy}, De Becker et al.\ \cite{cyg8a}, Naz\'e et al.\ \cite{cyg92}), whilst the fourth one (Cyg\,OB2 \#12) is a luminous blue variable (LBV) candidate and could be one of the most luminous and optically brightest stars in our Galaxy (e.g.\ Massey \& Thompson \cite{massey}). 

X-ray emission from a handful of massive stars in Cyg\,OB2 was discovered serendipitously when the {\it EINSTEIN} observatory was pointed at Cyg\,X-3 (Harnden et al.\ \cite{Harnden}). All subsequent major X-ray satellites have observed this association. The first CCD X-ray spectra of the four brightest massive stars (Cyg\,OB2 \#5, 8a, 9, and 12) in the association were obtained with {\it ASCA}-SIS (Kitamoto \& Mukai \cite{KM}). The {\it Chandra} observatory has observed Cyg\,OB2 several times, both with the HETG, to gather high-resolution X-ray spectroscopy of the brightest sources (Waldron et al.\ \cite{Waldron}), and the ACIS-I instrument, to collect CCD images and spectra of the members of the association (Albacete-Colombo et al.\ \cite{AC1}, \cite{AC2}, Wright \& Drake \cite{WD1}, Wright et al.\ \cite{WDDV})\footnote{Most recently Cyg\,OB2 has been the target of a {\it Chandra} legacy survey (Drake \cite{Drake}).}. Thanks to the very narrow PSF and the low background, the 100\,ksec ACIS-I exposure of the central region of Cyg\,OB2 (centered near Cyg\,OB2 \#9) revealed about 1300 sources (Wright \& Drake \cite{WD1}). Most of these sources are very faint and are likely associated with pre-main sequence stars. This is unsurprising as {\it IRAS} and $^{12}$CO surveys revealed the Cyg\,OB2 region to be a highly active star-forming region (Odenwald \& Schwartz \cite{OS}, Parthasarathy et al.\ \cite{Parta}). 

The distance to the Cyg\,OB2 association remains somewhat controversial. Massey \& Thompson (\cite{massey}) derived a distance modulus of $11.2 \pm 0.1$, in agreement with the more recent mean spectro-photometric distance obtained by Kiminki et al.\ (\cite{Kiminki}). Hanson (\cite{hanson}) used a cooler O-star temperature scale than Massey \& Thompson (\cite{massey}) and inferred a lower distance modulus of 10.4\,mag (implying luminosities that are lower by a factor 2). An even lower distance modulus (9.8\,mag) was obtained from the analysis of the light curve of the eclipsing binary Cyg\,OB2 \#5 (Linder et al.\ \cite{cyg5}), although this latter result might be biased by the presence of third light as Cyg\,OB2 \#5 most likely hosts four stars (Kennedy et al.\ \cite{Kennedy}). Throughout this paper, we adopt the 10.4\,mag distance-modulus of Hanson (\cite{hanson}), corresponding to a distance of 1.2\,kpc,\footnote{This value agrees with the distance estimate of $1.32^{+0.11}_{-0.09}$\,kpc towards the W75N star forming region (Rygl et al.\ \cite{Rygl}), which is part of the Cygnus X giant molecular cloud, as is Cyg\,OB2.} although we emphasize that most of our conclusions are essentially independent of the distance.

We report the results of our analysis of six {\it XMM-Newton} observations of the core of Cyg\,OB2. The main purpose of this campaign was to study the X-ray properties and variability of the X-ray bright massive members of the association. Large X-ray luminosities, X-ray variability and/or spectral hardness are frequently considered as indicators of either interacting wind binaries or single stars with magnetic fields (see e.g.\ Gagn\'e et al.\ \cite{Gagne}), although this is certainly not a one to one relationship and some O-star binaries actually display X-ray properties quite similar to those of normal single O-stars (e.g.\ Naz\'e et al.\ \cite{CCCP}). When interpreting our data, we thus have to keep in mind the latest results of multiplicity studies in Cyg\,OB2 (Kiminki et al.\ \cite{Kiminki09} and references therein).

This paper is outlined as follows. In Sect.\,\ref{observations}, we present our data set and the data reduction. The general properties of the X-ray and UV sources in our field of view as well as their long and short term variability are discussed in Sect.\,\ref{general}. Section \ref{specOB} presents a discussion of the X-ray emission of five relatively bright OB stars and Sect.\,\ref{conclusion} summarizes our conclusions.  

\section{{\it XMM-Newton} observations of Cyg\,OB2 \label{observations}}
Six pointings centred on Cyg\,OB2 \#8a were taken with the {\it XMM-Newton} satellite (Jansen et al.\ \cite{Jansen}). The first four observations were obtained in October -- November 2004 with roughly ten days between two consecutive pointings. Two follow-up observations were gathered in April -- May 2007 (see Table\,\ref{journal}). The EPIC cameras were used with the medium filter to reject optical light. 

\begin{figure}[thb]
\begin{center}
\resizebox{8cm}{!}{\includegraphics{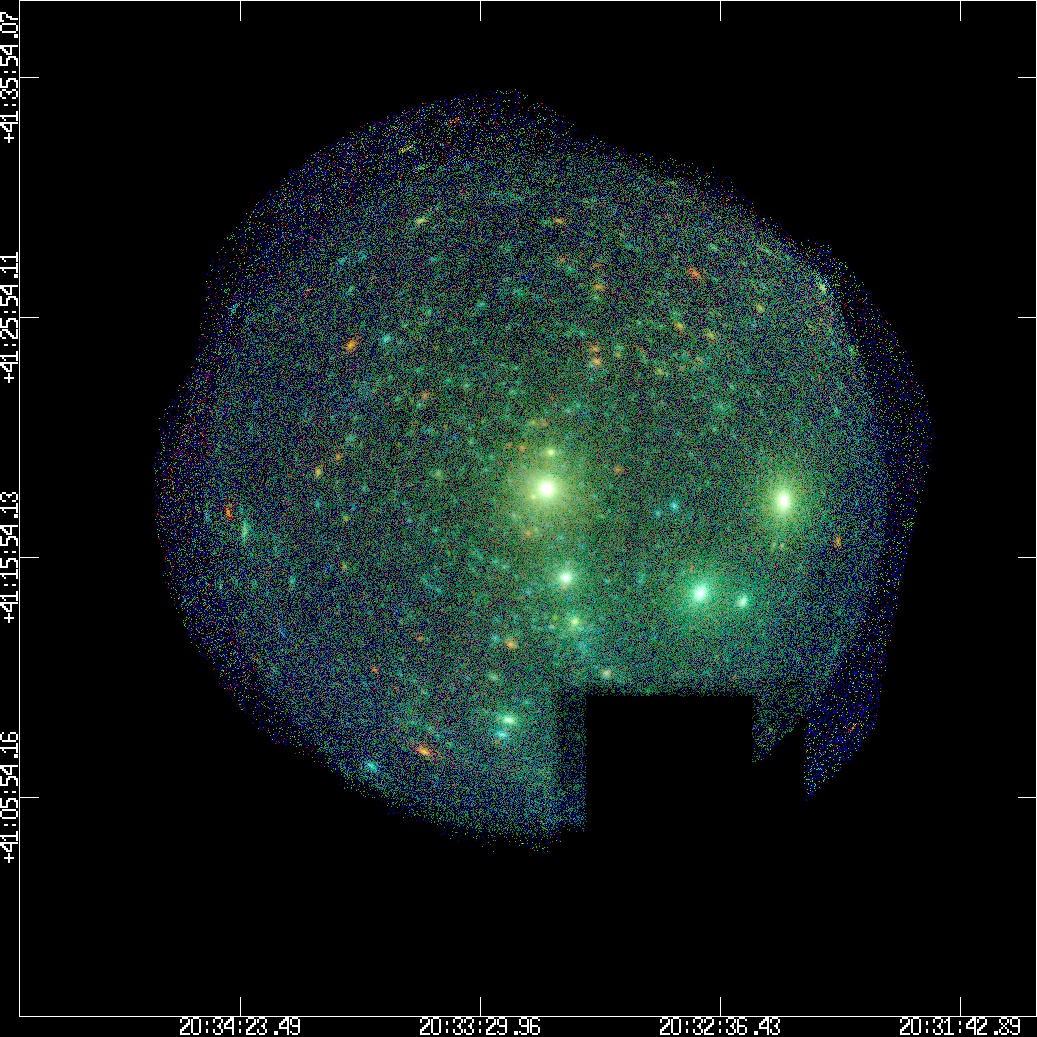}}
\end{center}
\caption{Energy-coded three-colour image built from our six {\it XMM-Newton} observations of Cyg\,OB2. The red, green, and blue colours correspond to the soft, medium, and hard energy bands used throughout this paper (see text). The individual EPIC images were exposure corrected, the stray-light features from Cyg\,X-3 (lower right corner) were excluded, and the out-of-time events from the bright central source (Cyg\,OB2 \#8a) were removed before the images were combined. A preliminary version of this figure was published by Rauw et al.\ (\cite{RNO}).\label{3colours}}
\end{figure}

The raw data were processed with SAS software version 10.0. Several observations (especially the last two) were affected by high background events (so-called soft-proton flares). These episodes were rejected in our processing and the remaining total useful exposure times were 105\,ksec for both EPIC-MOS (Turner et al.\ \cite{MOS}) detectors and 74\,ksec for the EPIC-pn (Str\"uder et al.\ \cite{pn}). Images were extracted for the soft (0.5 -- 1.0\,keV), medium (1.0 -- 2.0\,keV), and hard (2.0 -- 8.0\,keV) energy bands. 
A few stray-light features, caused by singly reflected photons from Cyg\,X-3, affect a small region in the lower right corner of the images. These parts of the images were excluded from the subsequent analysis (see Fig.\ \ref{3colours}). 

We note that the best-fit positions of TeV\,J2032+4130 are just outside our field of view (Aharonian et al.\ \cite{Felix}, Albert et al.\ \cite{Albert}). Apparently diffuse X-ray emission from this TeV source was discovered with another {\it XMM-Newton} observation (Horns et al.\ \cite{Horns}).   

\begin{table*}
\caption{Journal of the six {\it XMM-Newton} observations discussed in this paper.\label{journal}}
\begin{center}
\begin{tabular}{c c c c c c c c}
\hline
Obs \# & Date & \multicolumn{2}{c}{Exposure time} & \multicolumn{2}{c}{Clean time} & \multicolumn{2}{c}{OM exposure time} \\
& JD-2450000 & \multicolumn{2}{c}{(ksec)} & \multicolumn{2}{c}{(ksec)} & \multicolumn{2}{c}{(ksec)}\\
& & MOS & pn & MOS & pn & UVW1 & UVM2\\
\hline 
1 & 3308.579 & 19.7 & 18.0 & 19.1 & 15.4 & 3.2 & 8.0 \\
2 & 3318.558 & 21.7 & 20.0 & 21.5 & 17.6 & 4.2 & 9.0 \\
3 & 3328.543 & 23.7 & 22.0 & 23.0 & 19.3 & 5.0 & 10.0\\
4 & 3338.506 & 21.7 & 20.0 & 13.7 & 10.9 & 0.0 & 8.2 \\
5 & 4220.355 & 30.5 & 28.9 & 11.8 &  3.6 & 10.3 & 12.3\\
6 & 4224.170 & 31.6 & 30.0 & 15.9 &  7.0 & 9.9 & 13.7\\
\hline
\end{tabular}
\tablefoot{Columns 5 and 6 provide the useful exposure times after discarding time intervals of high background.}
\end{center}
\end{table*}
Some preliminary results for the first four {\it XMM-Newton} pointings were discussed by Rauw et al.\ (\cite{jenam}), whilst full analyses of the data for Cyg\,OB2 \#8a, Cyg\,OB2 \#5, and Cyg\,OB2 \#9 were presented by De Becker et al.\ (\cite{cyg8a1}), Linder et al.\ (\cite{cyg5}), Naz\'e et al.\ (\cite{cyg9}), and Blomme et al.\ (\cite{cyg8a2}). These three objects are not discussed any further in this paper. 

\begin{figure*}[thb]
\begin{minipage}{8cm}
\begin{center}
\resizebox{8cm}{!}{\includegraphics{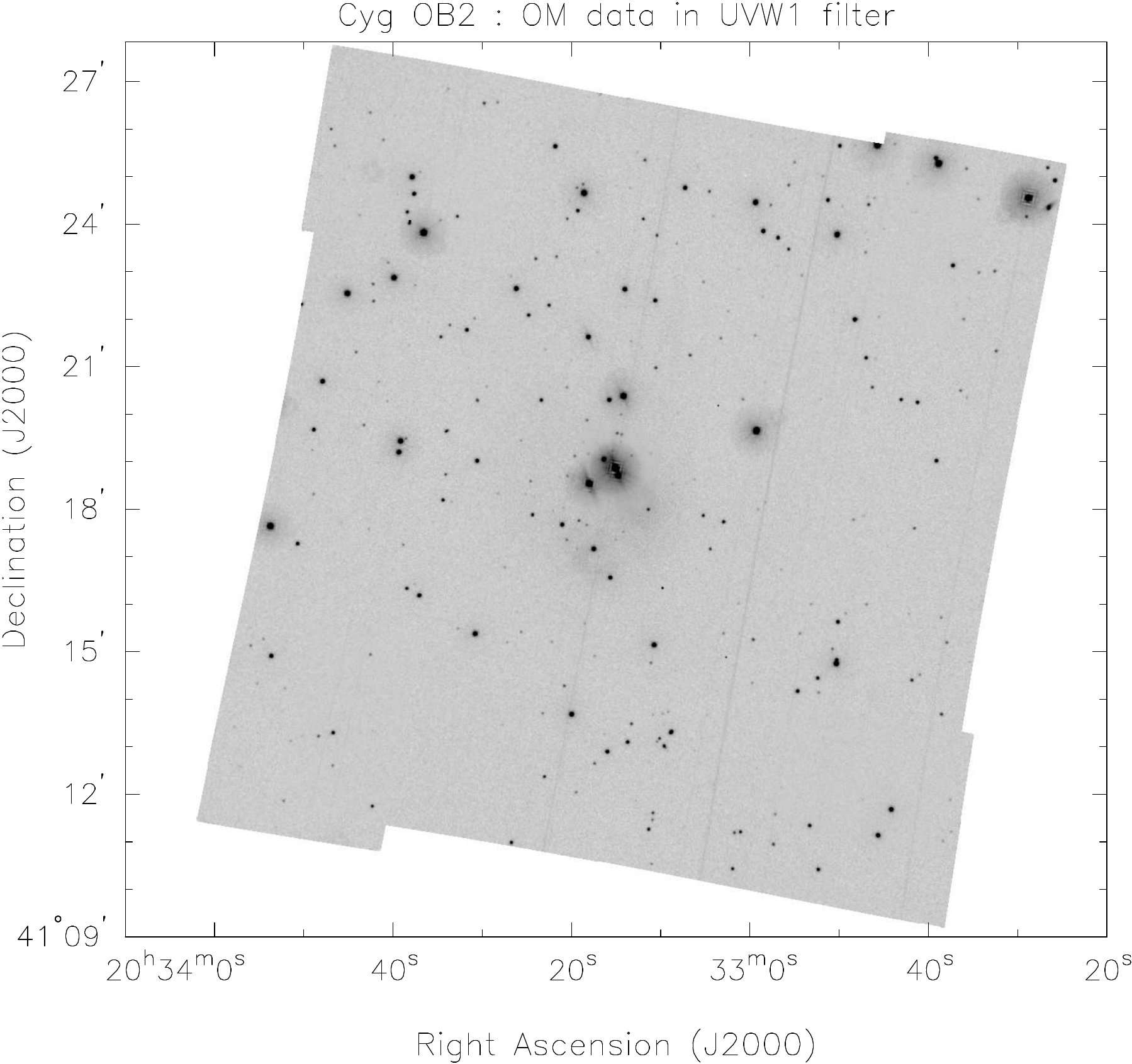}}
\end{center}
\end{minipage}
\hfill
\begin{minipage}{8cm}
\begin{center}
\resizebox{8cm}{!}{\includegraphics{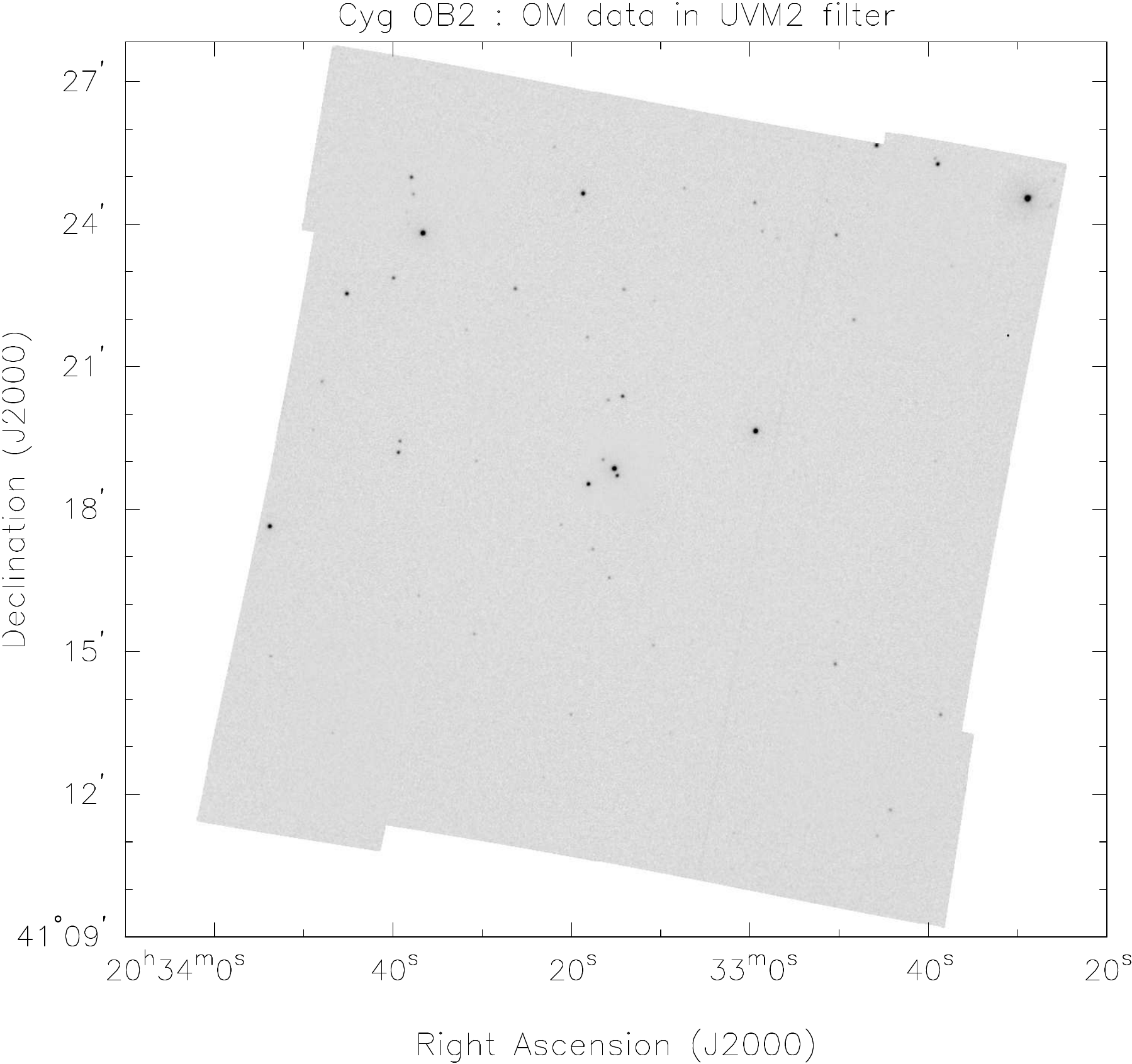}}
\end{center}
\end{minipage}
\caption{Mosaic image of the core of Cyg\,OB2 as seen with the Optical Monitor through the $UVW1$ (left) and $UVM2$ (right) filters during observation 5. The impact of the interstellar reddening on the number of sources detected in the $UVM2$ band is clearly seen. \label{OMimage}}
\end{figure*}

The Optical Monitor (OM, Mason et al.\ \cite{OM}) instrument onboard {\it XMM-Newton} observed Cyg\,OB2 through the $UVW1$ and $UVM2$ filters (see Fig.\ \ref{OMimage}). The first four pointings were performed in `full frame low-resolution' mode, whilst the last two used the `imaging mode' consisting of a mosaic of five exposures to cover the full 17\,arcmin square field of view (FoV) of the instrument (Fig.\ \ref{OMimage}). The data were processed with the relevant commands of the SAS software. 
The OM images display some artifacts such as smoke-ring ghosts caused by light reflected off a chamfer in the detector window and modulo-8 pattern around bright sources (see Mason et al.\ \cite{OM}). The OM detector has a read-out time of 11\,ms. The brightest sources in our field of view therefore suffer from coincidence losses that amount typically to 10\% for sources with 10\,counts\,s$^{-1}$ (Mason et al.\ \cite{OM}). These losses are partially recovered during the data processing to provide `corrected' count rates, which are then converted into AB magnitudes (Oke \cite{Oke}). In our cases, a count rate of 10\,counts\,s$^{-1}$ corresponds to $UVW1$ and $UVM2$ AB-magnitudes of 16.066 and 14.912. The astrometry of a dozen O-type stars, spread all over the OM field of view, was compared to the coordinates taken from the catalogue of Massey \& Thompson (\cite{massey}, hereafter MT91). We found a small shift (around 1.3\,arcsec in right ascension and 1.8\,arcsec in declination) for each observation. There was no obvious trend in this shift with position across the FoV, thus we applied a simple translation to the OM coordinates before performing cross-correlations with optical or X-ray catalogues. After this correction, the average distance between the OM sources and their counterpart in the MT91 catalogue amounts to 0.63\,arcsec for 225 objects (see Sect.\ \ref{general}). 

\section{The X-ray sources: general properties \label{general}}
A three colour X-ray image derived from our combined data set is shown in Fig.\,\ref{3colours}. It can be seen that the faint X-ray sources are found all over the field of view, although some areas contain larger concentrations of sources. Bica et al.\ (\cite{Bica}) reported the discovery of two open clusters in the core of Cyg\,OB2 roughly centered on stars \#8a and \#22, respectively. Whilst we find a relatively large number of X-ray sources within the regions defind by Bica et al.\ (\cite{Bica}), we note that there are other places inside our field of view with similar concentrations of X-ray sources (such as around 20h32m47s $+41^{\circ}$24'45").
\subsection{The nature of the X-ray sources}
The SAS routines detected a total of 206 sources in the combined EPIC images (outside the region affected by the straylight) with a significance threshold of 20\footnote{This implies a probability of $\leq e^{-20}$ that a random Poissonian fluctuation could have caused the observed source counts inside the detection cell. Over the entire EPIC field of view, the expected number of spurious detections amounts to about 2\% of the total number of detected sources.} and average count rates as low as 0.0004 and 0.0002\,cts\,s$^{-1}$ for EPIC-pn and EPIC-MOS. From a histogram of the number of detections versus count rate, we conclude that our detections are complete down to rates of 0.0022 and 0.0008\,cts\,s$^{-1}$ for EPIC-pn and EPIC-MOS, respectively. All these sources were verified by eye and 199 of them were eventually retained as real sources. There are a number of additional, fainter sources that were not selected by the detection algorithm, because they are located in crowded regions and their significance over the background therefore falls below our threshold. Finally, we stress that a few sources in our list could actually be blends of two fainter, spatially unresolved sources. Table\,\ref{newtable1} lists the coordinates of our sources, their X-ray properties, and the results of cross-correlation with various catalogues (see below). The full version of this table is available at the CDS. 

\begin{table*}
\caption{Properties of the X-ray sources in Cyg\,OB2 detected with {\it XMM-Newton}.\label{newtable1}}
\begin{tabular}{c c c c c c c c c c c c}
\hline
$[1]$ & $[2]$ & $[3]$ & $[4]$ & $[5]$ & $[6]$ & $[7]$ & $[8]$ & $[9]$ & $[10]$ & $[11]$ & $[12]$\\
\hline
  6 & 20:32:22.3 & +41:18:17.3 & 0.1584D+00 & 0.0162 & 6 & 0.6659D+00 & 0.0000 & 6 & 0.2717D+00 & 0.0000 & 6 \\ 
 13 & 20:32:31.4 & +41:14:06.0 & 0.6859D$-$02 & 0.0193 & 6 & 0.5691D$-$01 & 0.0000 & 6 & 0.4197D$-$01 & 0.0000 & 6 \\
 16 & 20:32:36.5 & +41:22:13.1 & 0.3771D$-$03 & 0.1853 & 4 & 0.2176D$-$02 & 0.0000 & 6 & 0.1183D$-$02 & 0.0000 & 5 \\
 22 & 20:32:40.8 & +41:14:28.0 & 0.3500D$-$01 & 0.0000 & 6 & 0.4164D+00 & 0.0000 & 6 & 0.3087D+00 & 0.0000 & 6 \\
 27 & 20:32:44.2 & +41:24:22.3 & 0.3257D$-$03 & 0.4902 & 4 & 0.1518D$-$02 & 0.0083 & 6 & 0.8545D$-$03 & 0.0017 & 5 \\
 34 & 20:32:46.6 & +41:18:06.8 & 0.2761D$-$03 & 0.0000 & 3 & 0.8789D$-$03 & 0.0000 & 3 & 0.6308D$-$03 & 0.0000 & 4 \\
 74 & 20:33:08.6 & +41:13:17.8 & 0.9955D$-$02 & 0.2646 & 5 & 0.5333D$-$01 & 0.0707 & 5 & 0.1364D$-$01 & 0.1639 & 5 \\
 80 & 20:33:10.6 & +41:15:07.6 & 0.4022D$-$01 & 0.1605 & 6 & 0.2614D+00 & 0.0005 & 6 & 0.1775D+00 & 0.1212 & 6 \\
 92 & 20:33:13.9 & +41:13:04.4 & 0.9154D$-$03 & 0.6731 & 6 & 0.4418D$-$02 & 0.0000 & 6 & 0.8938D$-$03 & 0.0000 & 6 \\
 95 & 20:33:13.9 & +41:20:21.1 & 0.1232D$-$01 & 0.0157 & 6 & 0.4790D$-$01 & 0.0814 & 6 & 0.9910D$-$02 & 0.1750 & 6 \\
 96 & 20:33:14.9 & +41:18:50.0 & 0.3530D+00 & 1.0000 & 6 & 0.1292D+01 & 0.0000 & 6 & 0.5778D+00 & 0.0000 & 6 \\
122 & 20:33:23.5 & +41:09:11.5 & 0.9762D$-$02 & 0.1457 & 6 & 0.7324D$-$01 & 0.0000 & 6 & 0.5054D$-$01 & 0.0457 & 6 \\
134 & 20:33:26.6 & +41:10:58.4 & 0.8776D$-$03 & 0.0082 & 5 & 0.5431D$-$02 & 0.0000 & 6 & 0.2440D$-$02 & 0.0000 & 6 \\ 
\hline
\end{tabular}
\vspace*{2mm}

\begin{tabular}{c c c c c c c c c c c c c}
\hline
$[1]$ & $[13]$ & $[14]$ & $[15]$ & $[16]$ & $[17]$ & $[18]$ & $[19]$ & $[20]$ & $[21]$ & $[22]$ & $[23]$ & $[24]$\\
\hline
  6 & 0.4100D$-$01 & 0.0001 & 6 & 0.2546D+00 & 0.0000 & 6 & 0.1081D+00 & 0.0000 & 6 & 0.4230D$-$01 & 0.0020 & 6 \\
 13 & 0.1747D$-$02 & 0.2114 & 6 & 0.2484D$-$01 & 0.0000 & 6 & 0.1794D$-$01 & 0.0000 & 6 & 0.1948D$-$02 & 0.0024 & 6 \\
 16 & 0.1344D$-$03 & 0.6146 & 4 & 0.7055D$-$03 & 0.0000 & 6 & 0.6065D$-$03 & 0.0000 & 5 & 0.8757D$-$04 & 0.6374 & 3 \\
 22 & 0.8892D$-$02 & 0.0032 & 6 & 0.1429D+00 & 0.0000 & 6 & 0.1087D+00 & 0.0000 & 6 & 0.7893D$-$02 & 0.0000 & 6 \\
 27 & 0.1216D$-$03 & 0.4100 & 2 & 0.5925D$-$03 & 0.3433 & 4 & 0.4596D$-$03 & 0.9076 & 3 & 0.1947D$-$03 & 0.9668 & 4 \\
 34 & 0.1740D$-$03 & 0.3820 & 3 & 0.7237D$-$02 & 0.0000 & 2 & 0.3863D$-$03 & 0.0000 & 4 & 0.1304D$-$03 & 0.0282 & 2 \\
 74 & 0.1779D$-$02 & 0.0166 & 6 & 0.1772D$-$01 & 0.1740 & 6 & 0.5399D$-$02 & 0.0019 & 6 & 0.1847D$-$02 & 0.1020 & 6 \\ 
 80 & 0.9531D$-$02 & 0.0093 & 6 & 0.8768D$-$01 & 0.0370 & 6 & 0.6220D$-$01 & 0.0007 & 6 & 0.9659D$-$02 & 0.0000 & 6 \\
 92 & 0.2085D$-$03 & 0.6109 & 5 & 0.7190D$-$03 & 0.0879 & 6 & 0.4920D$-$03 & 0.0000 & 5 & 0.2043D$-$03 & 0.8327 & 5 \\ 
 95 & 0.2614D$-$02 & 0.0900 & 6 & 0.1552D$-$01 & 0.2145 & 6 & 0.3727D$-$02 & 0.7770 & 6 & 0.2804D$-$02 & 0.1776 & 6 \\
 96 & 0.8258D$-$01 & 1.0000 & 6 & 0.4440D+00 & 0.0000 & 6 & 0.2159D+00 & 0.0000 & 6 & 0.8393D$-$01 & 1.0000 & 6 \\
122 & 0.2441D$-$02 & 0.2976 & 6 & 0.2709D$-$01 & 0.0004 & 6 & 0.1681D$-$01 & 0.0000 & 6 & 0.2605D$-$02 & 0.9902 & 6 \\
134 & 0.2774D$-$03 & 0.4752 & 6 & 0.1713D$-$02 & 0.0034 & 6 & 0.6320D$-$03 & 0.0000 & 5 & 0.2919D$-$03 & 0.7785 & 4 \\ 
\hline
\end{tabular}
\vspace*{2mm}

\begin{tabular}{c c c c c c c c c c c c c c}
\hline
$[1]$ & $[25]$ & $[26]$ & $[27]$ & $[28]$ & $[29]$ & $[30]$ & $[31]$ & $[32]$ & $[33]$ & $[34]$ & $[35]$ & $[36]$ & $[37]$ \\
\hline
  6 & 0.2758D+00 & 0.0000 & 6 & 0.1158D+00 & 0.0000 & 6 &        &        &        &  5.187 &  4.745 &  4.339 & AEA \\
 13 & 0.2564D$-$01 & 0.0000 & 6 & 0.1902D$-$01 & 0.0000 & 6 &        &        &     &  7.817 &  7.094 &  6.664 & AAA \\
 16 & 0.4226D$-$03 & 0.0000 & 6 & 0.6045D$-$03 & 0.0000 & 2 &        &        &      & 12.106 & 11.337 & 10.879 & AAA \\
 22 & 0.1431D+00 & 0.0000 & 6 & 0.1080D+00 & 0.0000 & 6 &        & 20.104 &  19 &  4.667 &  3.512 &  2.704 & DDD \\
 27 & 0.5709D$-$03 & 0.6657 & 5 & 0.1174D$-$02 & 0.0081 & 5 &        &        &     & 13.820 & 12.754 & 12.408 & AAA \\   
 34 & 0.4953D$-$03 & 0.0000 & 4 & 0.4651D$-$03 & 0.0000 & 6 &        &        &               & 16.337 & 14.868 & 14.474 & AAA \\     
 74 & 0.1710D$-$01 & 0.0000 & 6 & 0.5112D$-$02 & 0.0005 & 6 & 20.747 & 16.699 &  95 &  7.110 &  6.540 &  6.226 & AAA \\ 
 80 & 0.9184D$-$01 & 0.0000 & 6 & 0.6456D$-$01 & 0.0004 & 6 & 19.642 & 15.811 & 103 &   6.468 &  5.897 &  5.570 & AAA \\
 92 & 0.1364D$-$02 & 0.0001 & 6 & 0.1356D$-$02 & 0.0000 & 5 & 20.904 & 17.445 & 118 &  9.034 &  8.559 &  8.280 & AAA \\
 95 & 0.1608D$-$01 & 0.2545 & 6 & 0.3704D$-$02 & 0.5000 & 6 & 17.385 & 14.174 & 120 &  7.248 &  6.818 &  6.611 & AAA \\ 
 96 & 0.4518D+00 & 0.0000 & 6 & 0.2164D+00 & 0.0000 & 6 & 15.267 & 12.462 & 126 &  6.123 &  5.721 &  5.503 & AAA \\ 
122 & 0.2510D$-$01 & 0.0965 & 6 & 0.1775D$-$01 & 0.3997 & 6 &        &        &     &  7.025 &  6.380 &  6.050 & AAA \\
134 & 0.1488D$-$02 & 0.1372 & 6 & 0.9622D$-$03 & 0.0000 & 5 &        &        &     &  8.971 &  8.434 &  8.165 & AAA \\
\hline
\end{tabular}
\vspace*{2mm}

\begin{tabular}{c c c c c c c c c c c c c}
\hline
$[1]$ & $[38]$ & $[39]$ & $[40]$ & $[41]$ & $[42]$ \\
\hline
  6 &       &       &      & O6.5-7\,I + OB-Ofpe/WN9 & Cyg\,OB2 \#5  \\
 13 & 12.87 & 1.22  & 2.19 & O7.5\,Ibf               & CPR2002 A11   \\
 16 & 16.19 & 1.16  & 2.06 &                         & MT91 291      \\
 22 & 11.46 & 2.37  & 3.35 & B5\,Ie                  & Cyg\,OB2 \#12 \\
 27 &       &       &      &                         &               \\
 34 &       &       &      &                         &               \\
 74 & 11.55 & 0.74  & 2.04 & O4\,III(f)              & Cyg\,OB2 \#22 \\
 80 & 10.96 & 0.69  & 1.81 & O5\,If + O6-7           & Cyg\,OB2 \#9  \\
 92 & 12.92 & 0.54  & 1.81 & O8\,V                   & MT91 455      \\
 95 & 10.55 & 0.19  & 1.45 & O3\,If                  & Cyg\,OB2 \#7  \\ 
 96 &  9.06 & 0.08  & 1.30 & O5.5\,I(f) + O6         & Cyg\,OB2 \#8a \\
122 & 11.84 & 0.93  & 2.20 & O5.5\,V                 & MT91 516      \\
134 & 13.00 & 0.61  & 1.87 & O8.5\,V                 & MT91 534      \\
\hline
\end{tabular}
\tablefoot{This table will be made available in its entirety at CDS. Some excerpts are shown here for guidance. The columns have the following meaning: $[1]$ = source number; $[2]$ and $[3]$ = source coordinates (J2000.0); $[4]$ -- $[6]$ = mean count rate in EPIC-pn soft band, probability that the source be constant and number of observations of this source in this band with EPIC-pn; $[7]$ -- $[9]$ = same for the EPIC-pn medium band; $[10]$ -- $[12]$ = same for the EPIC-pn hard band; $[13]$ -- $[15]$ = same for the EPIC-MOS1 soft band;
$[16]$ -- $[18]$ = same for the EPIC-MOS1 medium band; $[19]$ -- $[21]$ = same for the EPIC-MOS1 hard band; $[22]$ -- $[24]$ = same for the EPIC-MOS2 soft band; $[25]$ -- $[27]$ = same for the EPIC-MOS2 medium band; $[28]$ -- $[30]$ =
same for the EPIC-MOS2 hard band; $[31]$ and $[32]$ = $UVM2$ and $UVW1$ magnitudes; $[33]$ = ID number of OM counterpart; $[34]$ -- $[36]$ = $J$, $H$, $K_s$ photometry from 2MASS; $[37]$ = 2MASS quality flag or "ncp", "2cp" "3cp" where 2cp (3cp) means there are 2 (resp.\ 3) 2MASS counterparts inside the 4 arcsec correlation radius whilst ncp means no 2MASS counterpart was found inside the correlation radius; $[38]$ -- $[40]$ = $V$, $U-B$, $B-V$ from Massey \& Thompson (\cite{massey}); $[41]$ and $[42]$ = spectral type and common identifier.}
\end{table*}

We cross-correlated our final list of X-ray sources with the 2MASS point source catalogue (Cutri et al.\ \cite{2mass}, Skrutskie et al.\,\cite{Skrutskie}). On the basis of the technique of Jeffries et al.\ (\cite{Jeff97}), the optimal cross-correlation radius, which offers the optimal compromise between obtaining the maximum number of true identifications and avoiding contamination by spurious coincidences, was established to be 4\,arcsec. For this correlation radius, we expect to find fewer than 5\% of false cross-identifications. We also correlated our list of sources with the $U\,B\,V$ photometric catalogue of Massey \& Thompson (\cite{massey}) and the list of OB star candidates provided by Com\'eron et al.\ (\cite{Comeron}). Twenty-four positive matches were found with the MT91 catalogue, to which we can add Cyg\,OB2 \#5, which is not included in the compilation of Massey \& Thompson (\cite{massey}). Three OB star candidates listed by Com\'eron et al.\ (\cite{Comeron}) fall within the field of view of our data. Two of them, CPR2002\,A11\footnote{At this stage, we stress that the star MT91\,267 listed by Massey \& Thompson (\cite{massey}) and object CPR2002\,A11 of Com\'eron et al.\ (\cite{Comeron}) refer apparently to the same star. Kiminki et al.\ (\cite{Kiminki}) classified MT91\,267 as a G-type star, whilst Negueruela et al.\ (\cite{Negu}) quoted an O7.5\,Ibf spectral type for A11. Re-examination of the Kiminki et al.\ (\cite{Kiminki}) Keck spectrum of MT91\,267 by Kobulnicky (2011, private communication) confirmed that this star is indeed an O-type star.} and A43 (= Cyg\,OB2 \#16 = MT91\,299), are likely counterparts to X-ray sources, whilst the third (CPR2002\,A17) is not detected. Furthermore, among the four Br$\gamma$ emitters (CPR2002\,B5, B8, B11, B13, in addition to Cyg\,OB2 \#9) in our field of view listed by Com\'eron et al.\ (\cite{Comeron}), only star B8 is detected, although at a rather large angular distance from the position of the X-ray source. We find that a total of 20 X-ray sources correspond to O-type stars, and 5 are associated with B-type stars, including the B supergiant and LBV candidate Cyg\,OB2 \#12. In addition to Cyg\,OB2 \#5, \#8a, and \#9, there are 5 OB stars (Cyg\,OB2 \#12, \#7, \#22, MT91\,516, and CPR2002\,A11) that are sufficiently bright and clearly separated from other bright sources for their spectra to be studied during the individual pointings. Table \ref{OBtable} summarizes the general properties of those X-ray sources that are associated with known OB stars. 

\begin{figure}[htb]
\begin{center}
\resizebox{8cm}{!}{\includegraphics{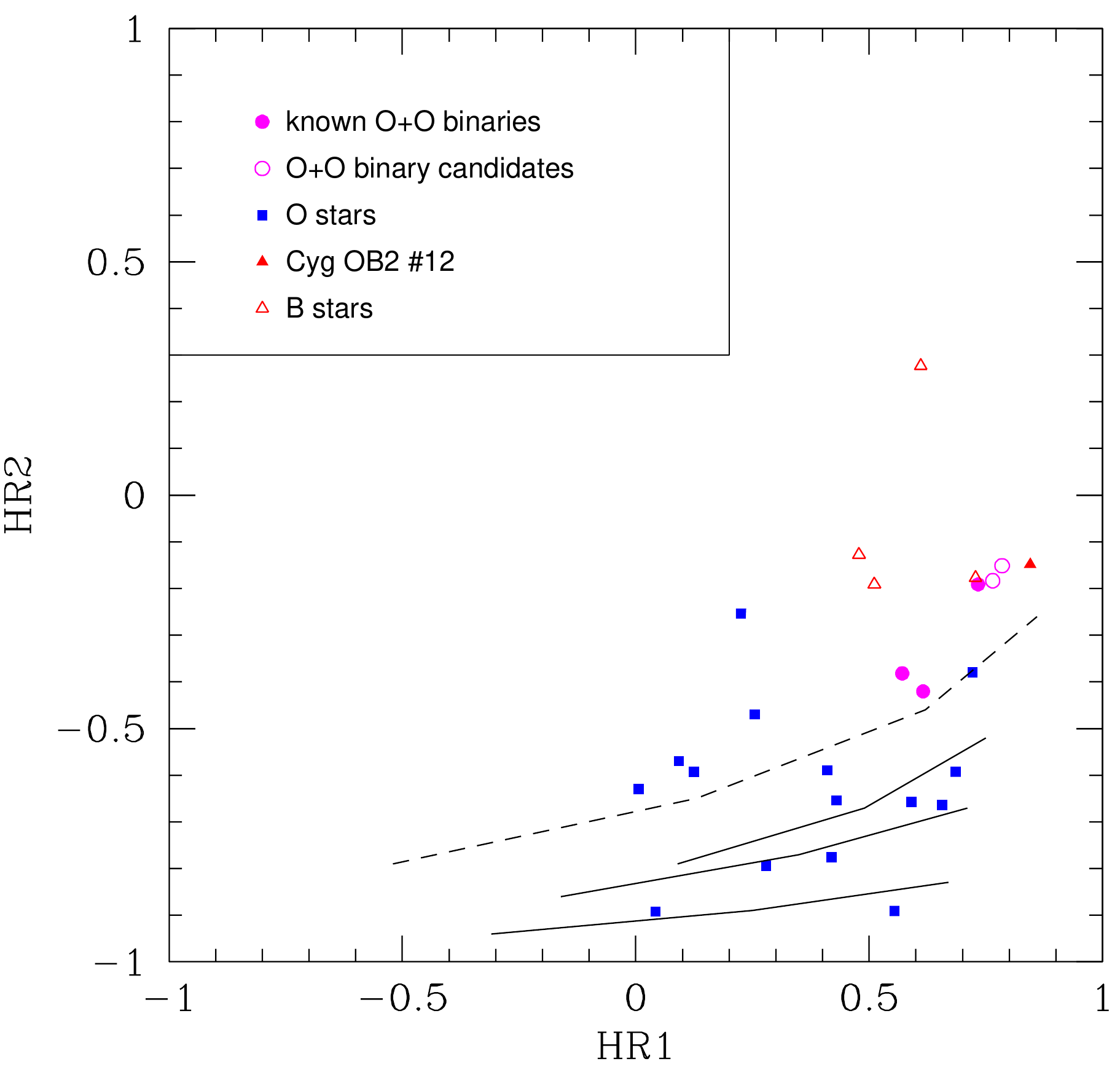}}
\end{center}
\caption{EPIC-pn hardness ratios of the OB stars in Cyg\,OB2. The solid lines yield the synthetic hardness ratios of absorbed single temperature thermal plasma models for temperatures of $kT$ = 0.4, 0.6, and 0.85\,keV (from bottom to top). The neutral column density increases along these lines from 0.5 to $1.5 \times 10^{22}$\,cm$^{-2}$ going from left to right. The dashed curve in turn indicates the hardness ratios of the 'typical' O-star two-temperature X-ray spectral model determined by Naz\'e (\cite{Naze}) with an absorption column-density increasing from 0.5 to $2.0 \times 10^{22}$\,cm$^{-2}$.\label{HRpn}}
\end{figure}

\begin{table*}[htb]
\caption{General properties of the OB stars in the core of Cyg\,OB2 as seen with {\it XMM-Newton}.\label{OBtable}}
\begin{tabular}{l l c c c c c c c c c c c c}
\hline
Name          & Spectral type           &  \multicolumn{4}{c}{pn}       && \multicolumn{2}{c}{MOS1} && \multicolumn{2}{c}{MOS2} & $UVM2$ & $UVW1$ \\
\cline{3-6}\cline{8-9}\cline{11-12}
              &                         &    CR  & HR1  &  HR2    & var &&  CR   & var &&  CR   & var &       &      \\
\hline 
\multicolumn{12}{l}{Objects studied in previous papers} & \\    
Cyg\,OB2 \#5  & O6.5-7\,I + OB-Ofpe/WN9 &  1.096 & 0.62 & $-0.42$ & SMH && 0.404 & SMH && 0.434 & SMH &       &       \\
Cyg\,OB2 \#8a & O5.5\,I(f) + O6         &  2.223 & 0.57 & $-0.38$ &  MH && 0.742 &  MH && 0.752 &  MH & 15.27 & 12.46 \\
Cyg\,OB2 \#9  & O5\,If + O6-7           &  0.479 & 0.73 & $-0.19$ &  M  && 0.159 & SMH && 0.166 & SMH & 19.64 & 15.81 \\
\vspace*{-3mm}\\
 \hline
\multicolumn{12}{l}{X-ray bright OB stars analysed in this paper} & \\ 
Cyg\,OB2 \#7  & O3\,If                  &  0.070 & 0.59 & $-0.66$ & S   && 0.022 &     && 0.023 &     & 17.39 & 14.17 \\ 
Cyg\,OB2 \#12 & B5\,Ie                  &  0.760 & 0.84 & $-0.15$ & SMH && 0.260 & SMH && 0.259 & SMH &       & 20.10 \\
Cyg\,OB2 \#22 & O4\,III(f)              &  0.077 & 0.69 & $-0.59$ &     && 0.025 & S H && 0.024 &  MH & 20.75 & 16.70 \\
MT91\,516     & O5.5\,V                 &  0.133 & 0.76 & $-0.18$ &  MH && 0.046 &  MH && 0.045 &     &       &       \\
CPR2002\,A11  & O7.5\,Ibf               &  0.106 & 0.78 & $-0.15$ & SMH && 0.044 &  MH && 0.047 & SMH &       &       \\
\vspace*{-3mm} \\
\hline
\multicolumn{12}{l}{X-ray fainter OB stars, not analysed in detail} & \\ 
Cyg\,OB2 \#4  & O7\,IIIf                &  0.018 & 0.23 & $-0.25$ &     && 0.011 &   H && 0.010 &     &       &       \\   
Cyg\,OB2 \#6  & O8\,V                   &  0.009 & 0.04 & $-0.89$ &     && 0.002 &     && 0.002 &     & 16.79 & 13.82 \\
Cyg\,OB2 \#8c & O4.5\,III(fc)           &  0.040 & 0.42 & $-0.78$ & S   && 0.014 &     && 0.014 &     & 16.58 & 13.41 \\
Cyg\,OB2 \#15 & O8\,V                   &  0.009 & 0.01 & $-0.63$ &  MH && 0.003 & S   && 0.003 &     &       &       \\
Cyg\,OB2 \#16 & O7\,V                   &  0.009 & 0.28 & $-0.79$ &     && 0.003 &     && 0.003 &     & 16.72 & 13.90 \\
Cyg\,OB2 \#17 & O8\,V                   &  0.005 & 0.09 & $-0.57$ &     && 0.002 &     && 0.002 &     & 18.18 & 15.13 \\
Cyg\,OB2 \#24 & O7\,V                   &  0.009 & 0.41 & $-0.59$ & S   && 0.002 &     && 0.003 &     & 19.39 & 15.97 \\
Cyg\,OB2 \#27 & O9.5\,V                 &  0.003 & 0.26 & $-0.47$ &   H && 0.001 &     && 0.001 &     & 19.82 & 16.45 \\
MT91\,311     & B2\,V                   &  0.003 & 0.61 &  0.28   &     && 0.001 &     && 0.001 &     & 20.94 & 18.01 \\
MT91\,322     & B2.5\,V                 &  0.002 & 0.48 & $-0.13$ &     && 0.001 &   H && 0.001 &     &       & 18.71 \\
MT91\,336     & B3\,III                 &  0.004 & 0.73 & $-0.18$ &     && 0.001 &     && 0.001 &     & 20.55 & 17.75 \\
MT91\,376     & O8\,V                   &  0.005 & 0.12 & $-0.59$ &     && 0.001 &     && 0.001 &     & 18.44 & 15.43 \\
MT91\,448     & O6\,V                   &  0.004 & 0.55 & $-0.89$ &     && 0.002 &     && 0.001 &  M  &       & 19.04 \\
MT91\,455     & O8\,V                   &  0.006 & 0.66 & $-0.66$ &  MH && 0.001 &   H && 0.003 &  MH & 20.90 & 17.45 \\
MT91\,485     & O8\,V                   &  0.007 & 0.43 & $-0.65$ &     && 0.002 &     && 0.003 &     & 19.25 & 16.03 \\
MT91\,534     & O8.5\,V                 &  0.009 & 0.72 & $-0.38$ & SMH && 0.003 &  MH && 0.003 &   H &       &       \\
MT91\,573     & B3\,I                   &  0.002 & 0.51 & $-0.19$ &  MH &&       &     && 0.001 &     &       & 18.12 \\
\hline
\end{tabular}
\tablefoot{Columns 3, 7, and 9 provide the average count rates (cts\,s$^{-1}$) over the 0.5 -- 8.0\,keV energy band for the pn, MOS1, and MOS2 detectors, respectively. Columns 6, 8, and 10 indicate the energy bands for which the null hypothesis of a constant count rate is rejected at the 5\% level for the pn, MOS1, and MOS2 detectors. The hardness ratios (quoted in columns 4 and 5 for the pn instrument only) are defined as HR1 = (M$-$S)/(M+S) and HR2 = (H$-$M)/(H+M). The X-ray spectra of the first three objects were discussed in previous papers, those of the five following objects are discussed in detail in this paper and the remainder of the objects are too faint for spectral analysis.}
\end{table*}

Figure\,\ref{HRpn} displays the hardness ratios HR2 = (H$-$M)/(H+M) versus HR1 = (M$-$S)/(M+S) for the OB-stars, where S, M, and H stand for the EPIC-pn count rates in the soft, medium, and hard energy bands, respectively. The solid lines yield the expected hardness ratios for single temperature plasma emission at $kT$ = 0.4, 0.6, and 0.85\,keV for neutral hydrogen column densities between 0.5 and $1.5 \times 10^{22}$\,cm$^{-2}$. The dashed line yields the average O-star X-ray spectrum as inferred by Naz\'e (\cite{Naze}) and absorbed by column densities ranging between 0.5 and $2.0 \times10^{22}$\,cm$^{-2}$. This latter model features two thermal plasma components with $kT_1$ and $kT_2$ equal to 0.24 and 2.3\,keV, respectively, and the hotter component contributing 16\% of the emission at 1\,keV. As can be seen from this figure, with a few exceptions, the hardness ratios of the majority of the O-type stars in Cyg\,OB2 can be understood in terms of typical O-star emission spectra affected by a large interstellar absorption. There is no evidence that the majority of the O-type stars have intrinsically hard spectra. 
The hardest X-ray emission is observed for the B-type stars. However, it remains unclear whether these B-stars (except for Cyg\,OB2 \#12) are intrinsic X-ray emitters. Their emission might also originate from hidden low-mass pre-main-sequence companions (e.g. Evans et al.\ \cite{Evans} and references therein), although we stress that in our dataset the B stars display little evidence of X-ray variability that would hint at the presence of a low-mass companion.

\begin{figure*}[t!hb]
\begin{minipage}{8cm}
\begin{center}
\resizebox{8cm}{!}{\includegraphics{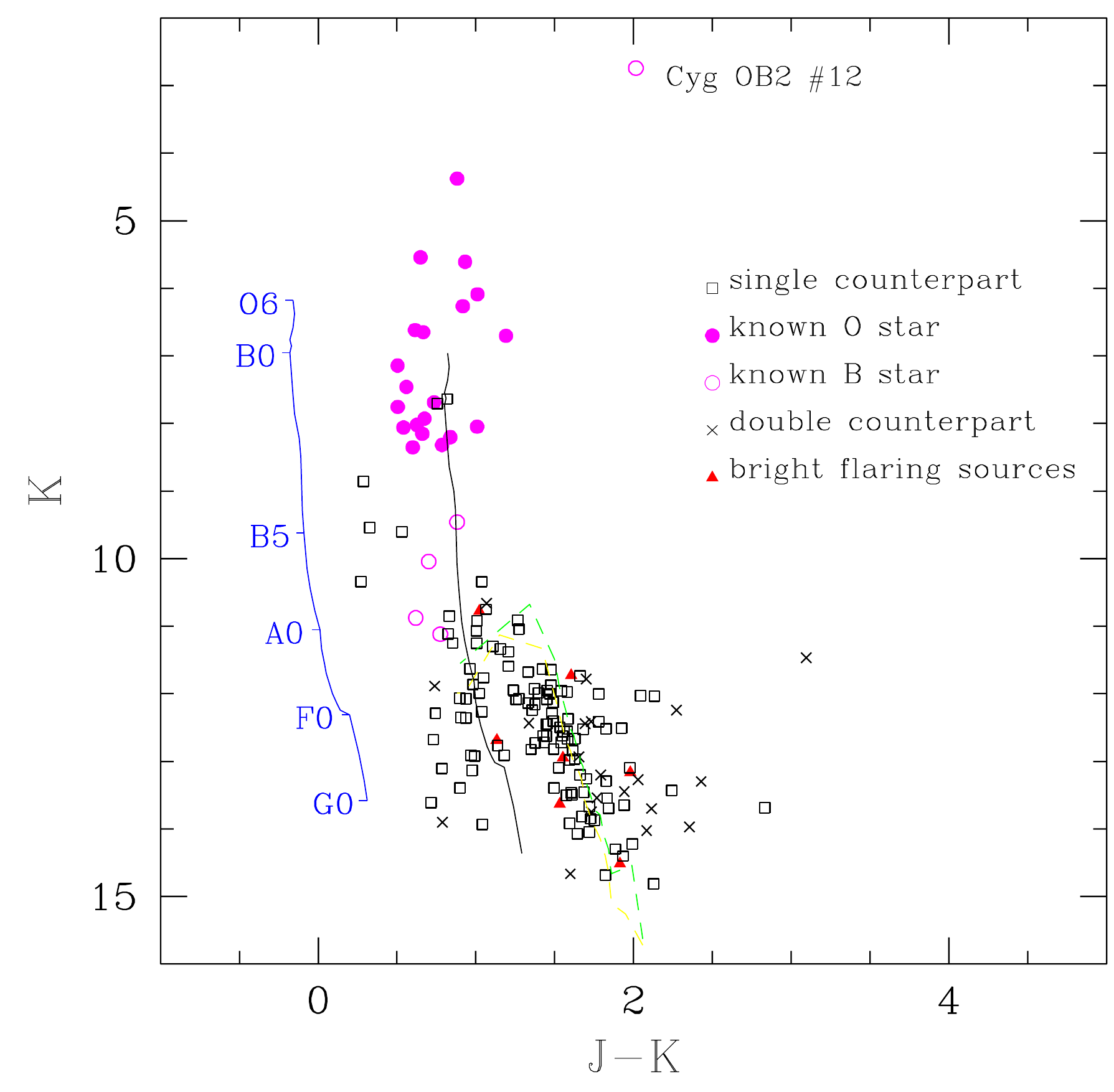}}
\end{center}
\end{minipage}
\hfill
\begin{minipage}{8cm}
\begin{center}
\resizebox{8cm}{!}{\includegraphics{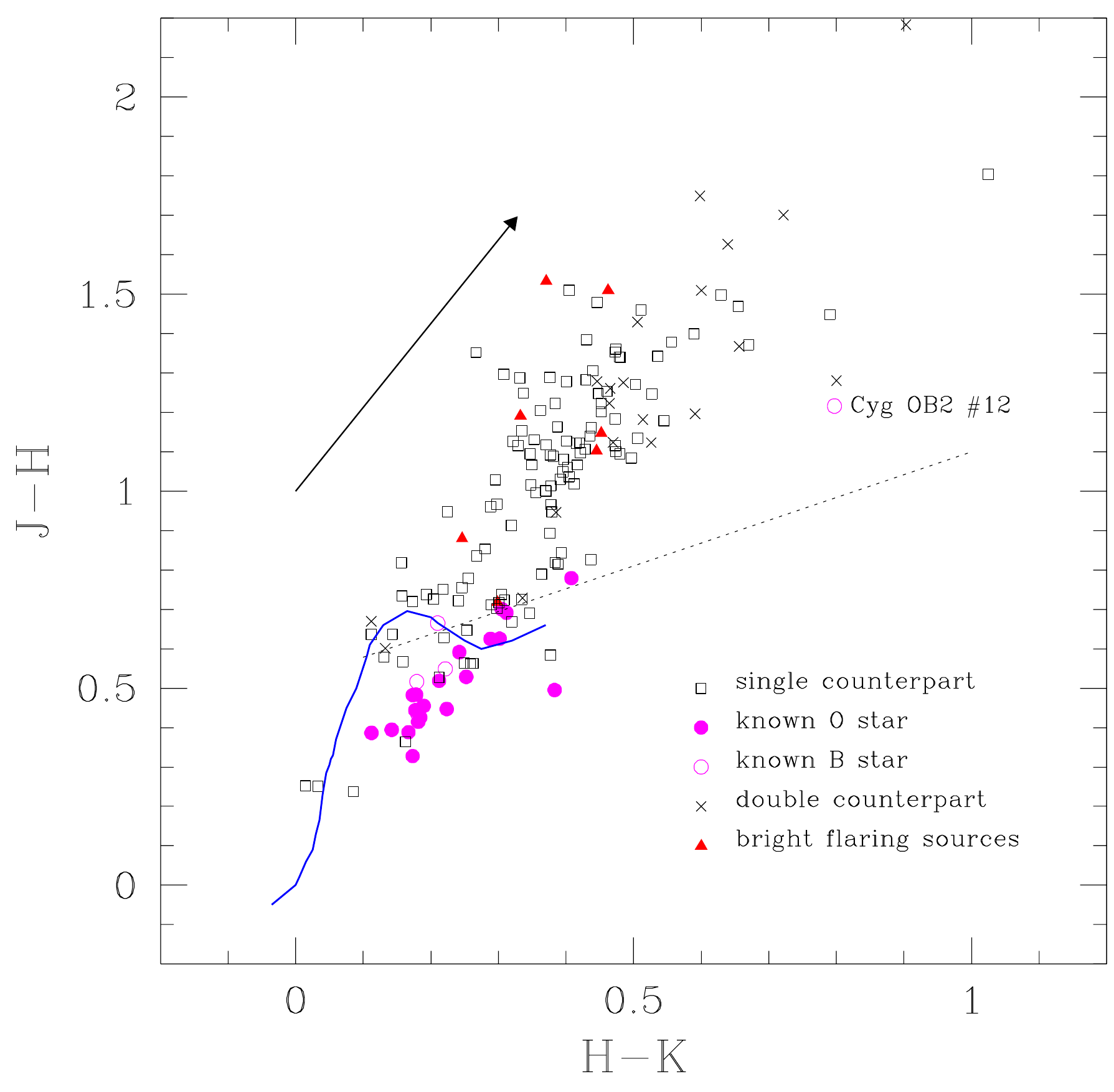}}
\end{center}
\end{minipage}
\caption{Left panel: $(J - K, K)$ near-IR colour-magnitude diagram of the 2MASS counterparts of our X-ray sources. The main-sequence (from Kn\"odlseder \cite{knoedlseder}) is shown for a distance modulus of 10.4 and for two different values of the reddening (from left to right, $A_{K} = 0.0$ and 0.78, Bica et al.\ \cite{Bica}). The reddened pre-main sequence isochrones from Siess et al.\ (\cite{Siess}) are shown for ages of 2 and 3 Myrs (dashed lines). Right panel: $(H - K, J -H)$ colour-colour diagram of the 2MASS counterparts. The solid lines yield the locus of the main-sequence according to Bessell \& Brett (\cite{BB}), while the dashed straight line indicates the locus of the unreddened classical T-Tauri stars following Meyer et al.\ (\cite{Meyer}). The reddening vector is shown according to Bica et al.\ (\cite{Bica}). \label{2mass}}
\end{figure*}

From the results of the cross-correlation with the 2MASS data, we compiled near-IR colour-magnitude and colour-colour diagrams. For this purpose, we used the colour transformations of Carpenter (\cite{Carpenter}, updated in 2003\footnote{http://www.ipac.caltech.edu/2mass/index.html}) to convert the 2MASS colours into the homogenized $J\,H\,K$ photometric system of Bessell \& Brett (\cite{BB}). Figure\,\ref{2mass} illustrates the results. The majority of the massive stars that are X-ray sources (except for Cyg\,OB2 \#12) have near-IR colours and magnitudes that are consistent with a reddening of $A_K\sim$ 0.5 -- 0.8 (see left panel of Fig.\,\ref{2mass}). The locus of the fainter counterparts in this diagram is considerably shifted to the red, which implies that these stars are on pre-main-sequence (PMS) tracks. 
\begin{figure*}[htb]
\begin{minipage}{8cm}
\begin{center}
\resizebox{8cm}{!}{\includegraphics{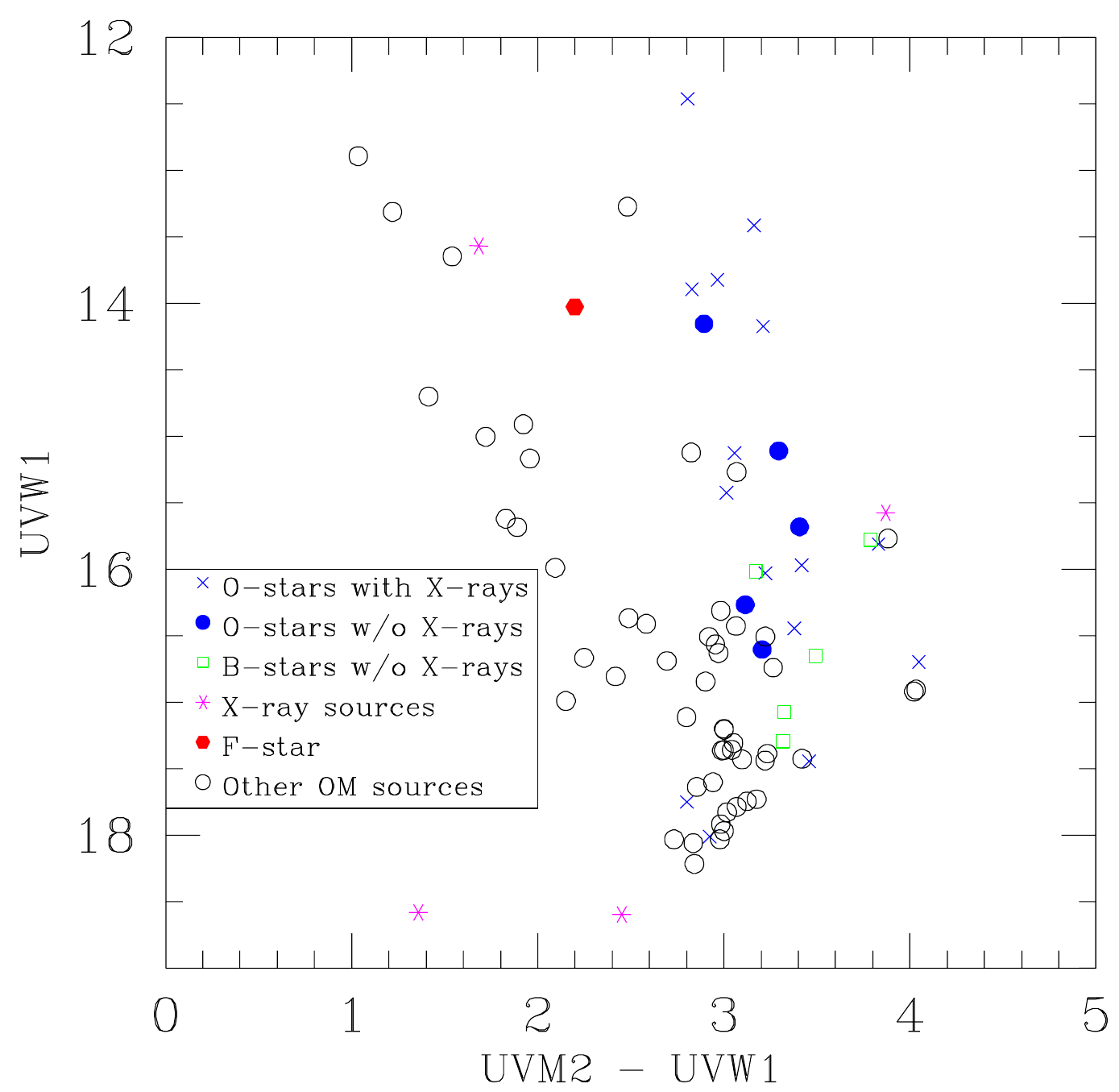}}
\end{center}
\end{minipage}
\hfill
\begin{minipage}{8cm}
\begin{center}
\resizebox{8cm}{!}{\includegraphics{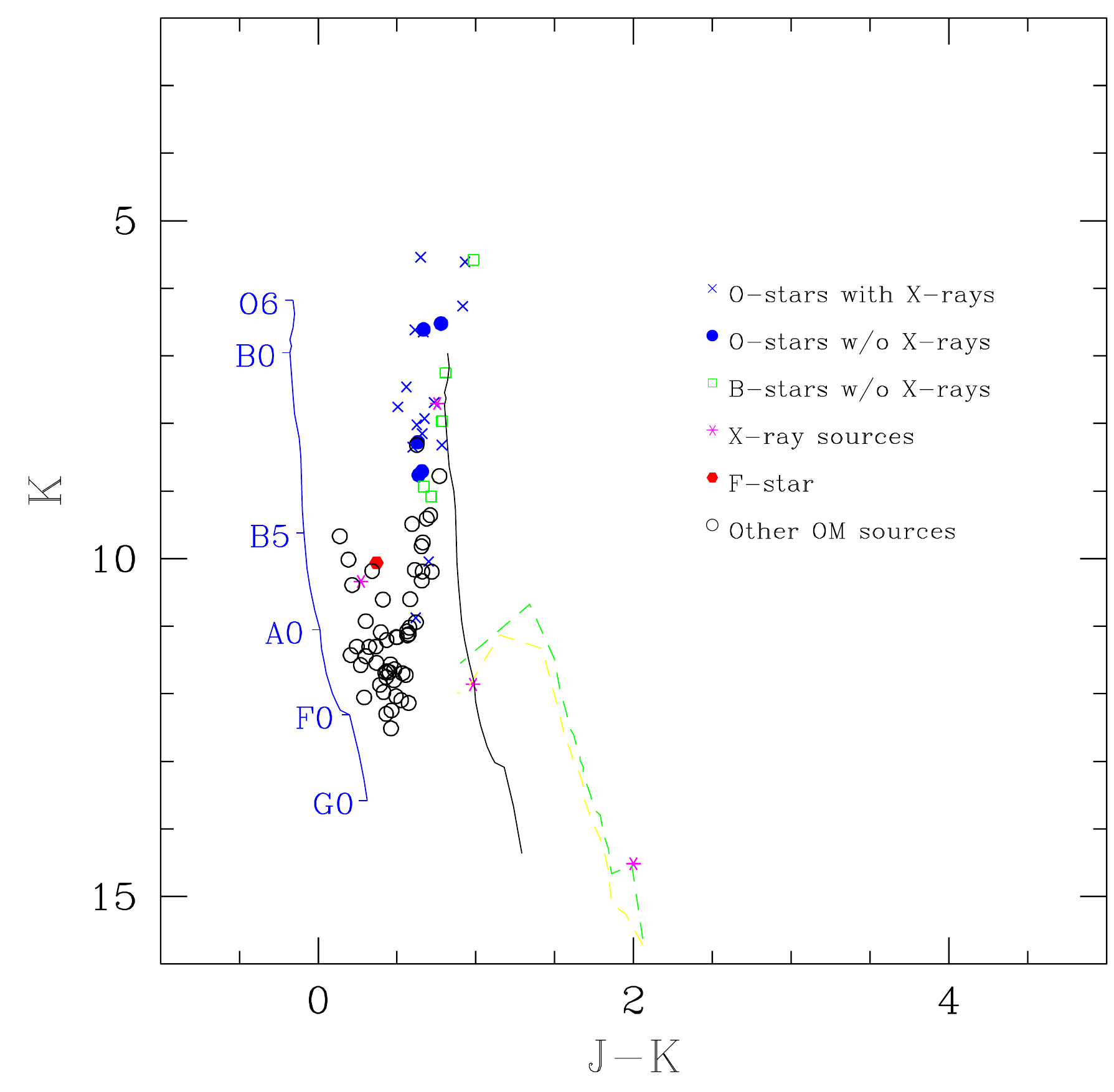}}
\end{center}
\end{minipage}
\caption{Left panel: colour-magnitude diagram compiled from the OM photometry of Cyg\,OB2. Right panel: $(J - K, K)$ near-IR colour-magnitude diagram (as Fig.\,\ref{2mass}) of the 2MASS counterparts of the OM sources that are detected in both UV filters. The F-type star indicated in these figures is Cyg\,OB2 \#72 (= MT91\,664, Massey \& Thompson \cite{massey}).\label{OMcoul}}
\end{figure*}
Bica et al.\ (\cite{Bica}) reported average values of $E(H - K_s) = 0.31$ and A$_{K_s}$ = 0.78 for the two clusters they identified at the core of Cyg\,OB2. The majority of the near-IR counterparts of our X-ray sources are consistent with this value of reddening and suggest an age of 2 -- 3\,Myr when compared to the PMS isochrones of Siess et al.\ (\cite{Siess}). These results are in excellent agreement with the conclusions of Albacete-Colombo et al.\ (\cite{AC1}) and Wright \& Drake (\cite{WD1}), although Wright et al.\ (\cite{WDDV}) inferred a slightly older age of $3.5_{-1.00}^{+0.75}$\,Myr. We draw attention to a group of sources that fall roughly along the main-sequence relation (shifted to the distance of Cyg\,OB2 and reddened according to the Bica et al.\ \cite{Bica} results), but below the main-sequence turn-on. One of these sources (number 86) displays a decaying flare during observation 1. These objects are likely to be late-type active foreground stars unrelated to Cyg\,OB2. 
From the right panel of Fig.\,\ref{2mass}, we see that there is little evidence of large IR excesses as one would expect for classical T\,Tauri stars (see Meyer et al.\ \cite{Meyer}). This is in line with the conclusion of Albacete-Colombo et al.\ (\cite{AC1}) and Wright et al.\ (\cite{WDDV}), who found that only 4 -- 8\% of the near-IR counterparts of their X-ray sources display infrared excesses that could hint at the presence of an accretion disk.

Vink et al.\ (\cite{Vink}) performed a search for H$\alpha$ emitters in Cyg\,OB2. Their field C overlaps to a large extent with our EPIC field of view. Four H$\alpha$ emitters tabulated by Vink et al.\ (\cite{Vink}) lie inside the EPIC FoV. Only one of them is marginally consistent with a detection in our X-ray data (source C9)\footnote{We note here that the position of source C9 in Fig.\ 1 of Vink et al.\ (\cite{Vink}) is inconsistent with the coordinates given in their Table\,2. We used the latter in our cross-correlation.}, although we do not regard this correlation as a significant detection.

Using our OM data, we also performed a search for UV sources. After rejecting some spurious sources (mainly due to the artifacts mentioned in Sect.\ \ref{observations}), we found a list of 254 objects that are detected at least in the $UVW1$ filter and at least during two different observations. Using a cross-correlation radius of 4\,arcsec, we find that 32 X-ray sources have a counterpart in our list of OM sources. Out of these 32 objects, 19 are measured in both OM filters, whilst the others are too faint in the $UVM2$ filter to be detected. Among the X-ray sources detected in both UV filters, 15 have an OB-type counterpart. Six X-ray flaring sources fall inside the OM FoV. Only one of them (the X-ray source 139) is detected with a magnitude of 18.8 in $UVW1$ (no detection in $UVM2$). From a cross-correlation with the catalogue of Massey \& Thompson (\cite{massey}), we find counterparts for 225 of our OM sources within a correlation radius of 2.0\,arcsec. Adopting the same correlation radius, all but five OM sources are found to have a 2MASS counterpart. The results of this correlation, along with the results of a $\chi^2$ variability test, are provided in Table 4 that is available at the CDS. 

The $UVM2$ filter contains the broad interstellar absorption feature at 2200\,\AA\ and is thus quite sensitive to the interstellar column density (Royer et al.\ \cite{Royer}). Given the rather heavy reddening towards the Cyg\,OB2 association, objects that are detected in both filters correspond to either the brightest and hottest association members or foreground stars. In the UV colour-magnitude diagram (see Fig.\,\ref{OMcoul}), the O-type stars define a clear, almost vertical sequence around $UVM2 - UVW1 \simeq 3$. The latter value is consistent with the large reddening towards the Cyg\,OB2 association (Royer et al.\ \cite{Royer}).    
  
For the O-type stars in our OM FoV, we inspected the inter-pointing light curves of these stars that were found to be variable using a $\chi^2$ test whilst not exceeding the limiting brightness at which 10\% coincidence losses occurr. Among the stars that did not display significant UV magnitude variations, the most remarkable cases are Cyg\,OB2 \#7, 9, and 12. The O-type stars that were found to be variable in either of the $UVM2$ and $UVW1$ bands are, respectively, Cyg\,OB2 \#8a, 8c, 16, 27 and Cyg\,OB2 \#22, 23, 27, MT91\,421, and 507. However, in most cases the peak-to-peak amplitude of this variability is limited to less than 0.15\,mag. Whilst this is larger than the formal errors in the magnitudes, the problems caused by the straylight features probably account for part of this apparent variability. There is one exception though: Cyg\,OB2 \#27 (Table\,\ref{S27}). The latter is a W UMa contact binary consisting of a late O and an early B-type star in a 1.46\,day orbit (Rios \& DeGioia-Eastwood \cite{Rios}). Unfortunately, there are no ephemerides available for this system, but we suspect that the 0.64\,mag drop that we observe in the $UVM2$ filter is likely due to the eclipse of the O-type star by its cooler companion. 
\addtocounter{table}{1}
\begin{table}[htb]
\caption{OM light curve of Cyg\,OB2 \#27 (= MT91\,696). \label{S27}}
\begin{center}
\begin{tabular}{c c c c c}
\hline
JD-2450000 & AB-magnitude \\
\hline
\multicolumn{2}{c}{$UVW1$}\\
$3308.5052 \pm 0.0184$ & $16.38 \pm 0.01$\\
$3318.4781 \pm 0.0242$ & $16.41 \pm 0.01$\\
$3328.4983 \pm 0.0289$ & $16.55 \pm 0.01$\\
\hline
\multicolumn{2}{c}{$UVM2$}\\
$3308.6205 \pm 0.0463$ & $19.86 \pm 0.11$\\
$3318.6049 \pm 0.0521$ & $19.63 \pm 0.08$\\
$3328.5934 \pm 0.0579$ & $20.27 \pm 0.13$\\
$3338.5573 \pm 0.0473$ & $19.78 \pm 0.08$\\
\hline
\end{tabular}
\tablefoot{The quoted uncertainty in the time corresponds to half of the exposure time.}
\end{center}
\end{table}
\begin{figure}[htb]
\begin{center}
\resizebox{8cm}{!}{\includegraphics{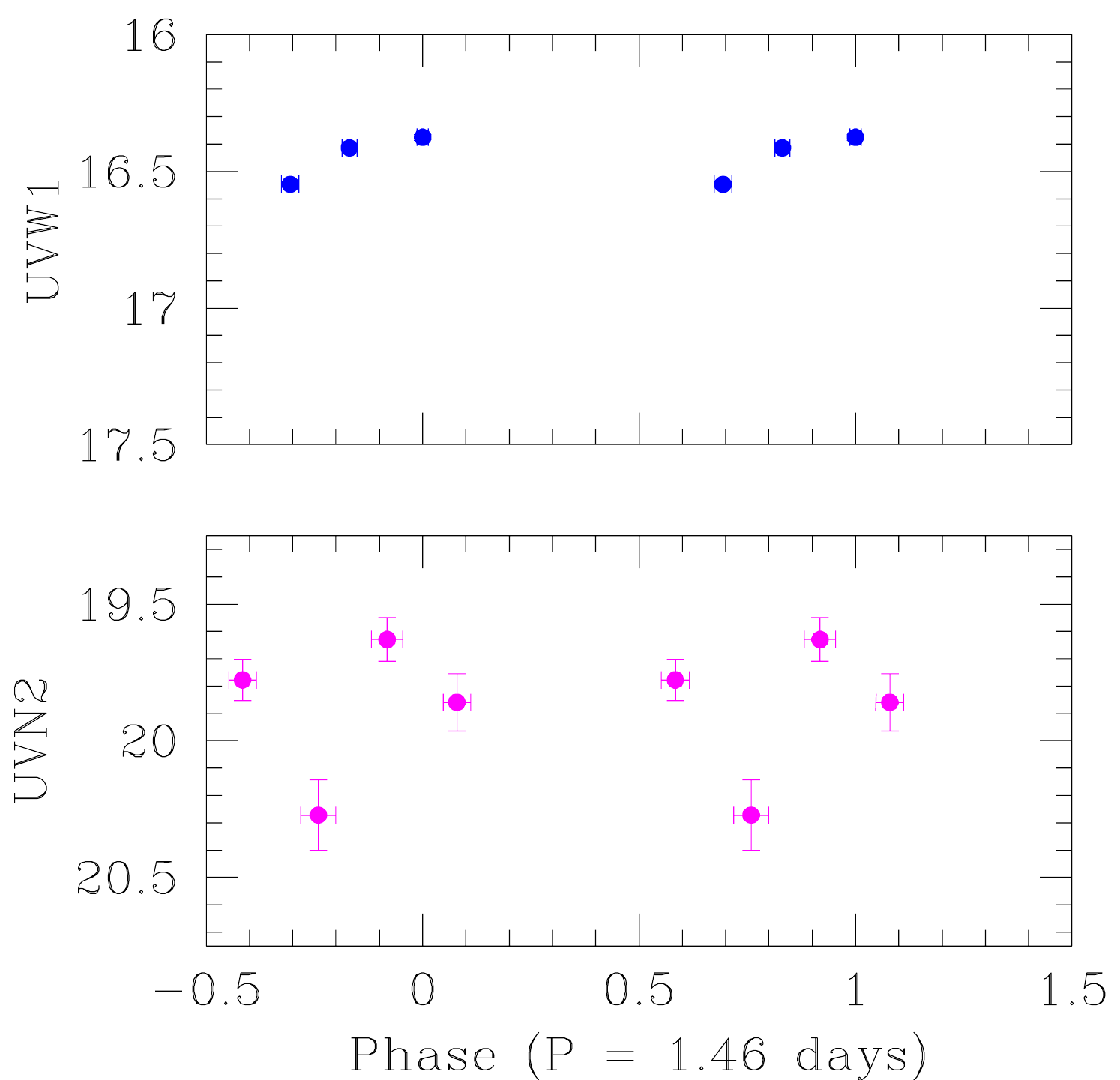}}
\end{center}
\caption{OM light curves of Cyg\,OB2 \#27 as a function of orbital phase. Phase 0.0 was arbitrarily chosen at the time of the first observation.\label{OMS27}}
\end{figure}

\subsection{Long and short term variations \label{lc}}
Using a $\chi^2$ test, we searched for inter-pointing variability. The variability test is performed using the mean exposure-corrected EPIC count rates from the different pointings, along with their estimated errors. The results are summarized in Table\,\ref{variability} for an adopted significance level of 95\%. As can be seen from that table, the incidence of variability increases significantly from the soft to the hard energy band. There are two reasons for this trend: (1) the huge interstellar absorption leads to rather low count rates in the soft energy band; and (2) a fraction of the variable sources display flares (see below) associated with a hardening of their spectra, hence the largest variability occurs in the hard energy band.

\begin{table}[htb]
\caption{Statistics of our inter-pointing $\chi^2$ variability test as a function of instrument and energy band.\label{variability}} 
\begin{center}
\begin{tabular}{c c c c c}
\hline
Instrument & Soft & Medium & Hard & Full \\
\hline
MOS1 &  5 & 29 & 39 & 52\\
MOS2 &  5 & 27 & 39 & 48\\
pn   & 14 & 43 & 47 & 69\\
\hline
\end{tabular}
\end{center}
\tablefoot{The numbers provide the fraction (in \%) of sources that display variability at a significance level of $\geq 95$\%.}
\end{table}

\begin{figure*}[h!tb]
\begin{minipage}{5.6cm}
\begin{center}
\resizebox{5.6cm}{!}{\includegraphics{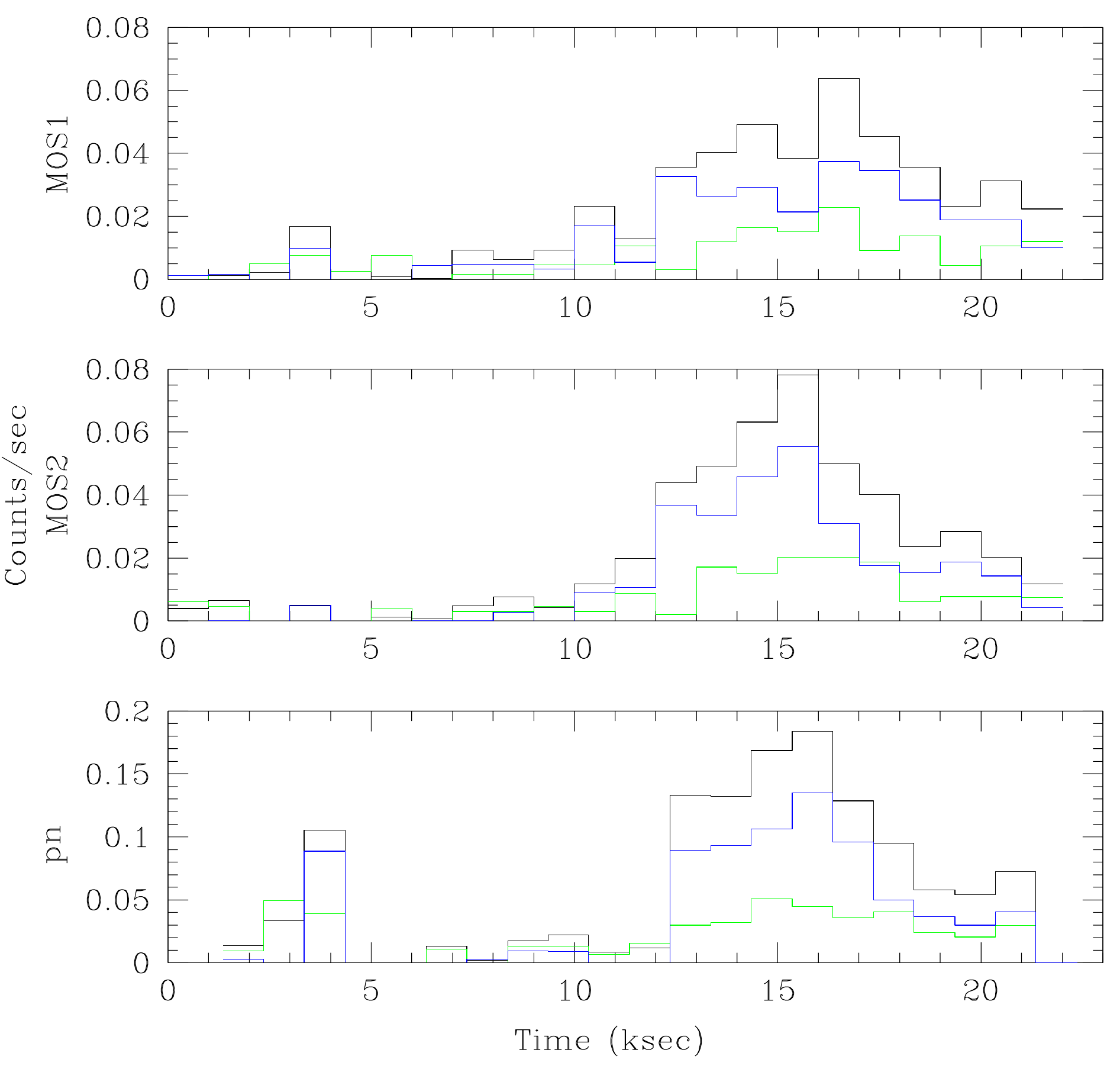}}
\end{center}
\end{minipage}
\hfill
\begin{minipage}{5.6cm}
\begin{center}
\resizebox{5.6cm}{!}{\includegraphics{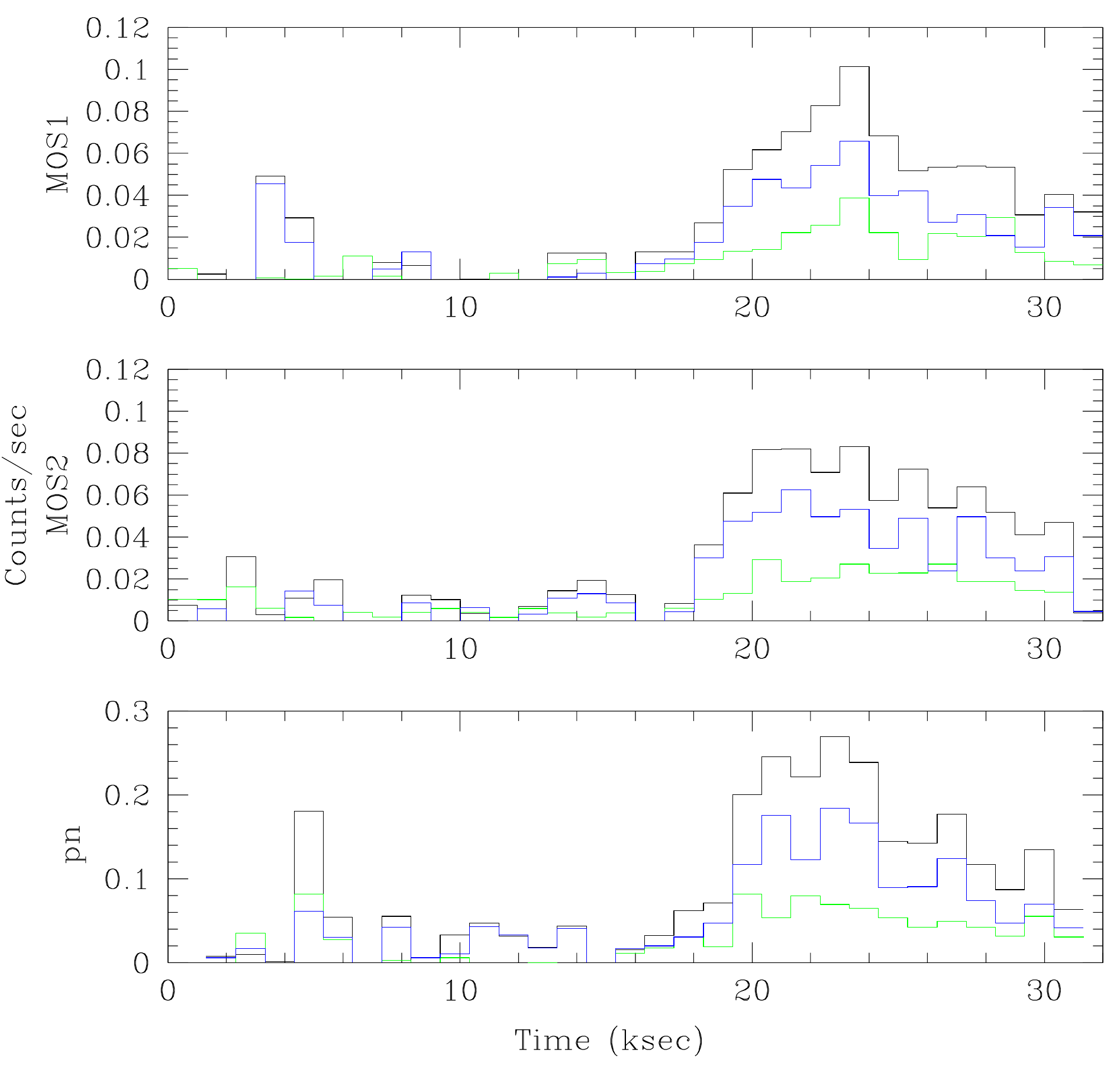}}
\end{center}
\end{minipage}
\hfill
\begin{minipage}{5.6cm}
\begin{center}
\resizebox{5.6cm}{!}{\includegraphics{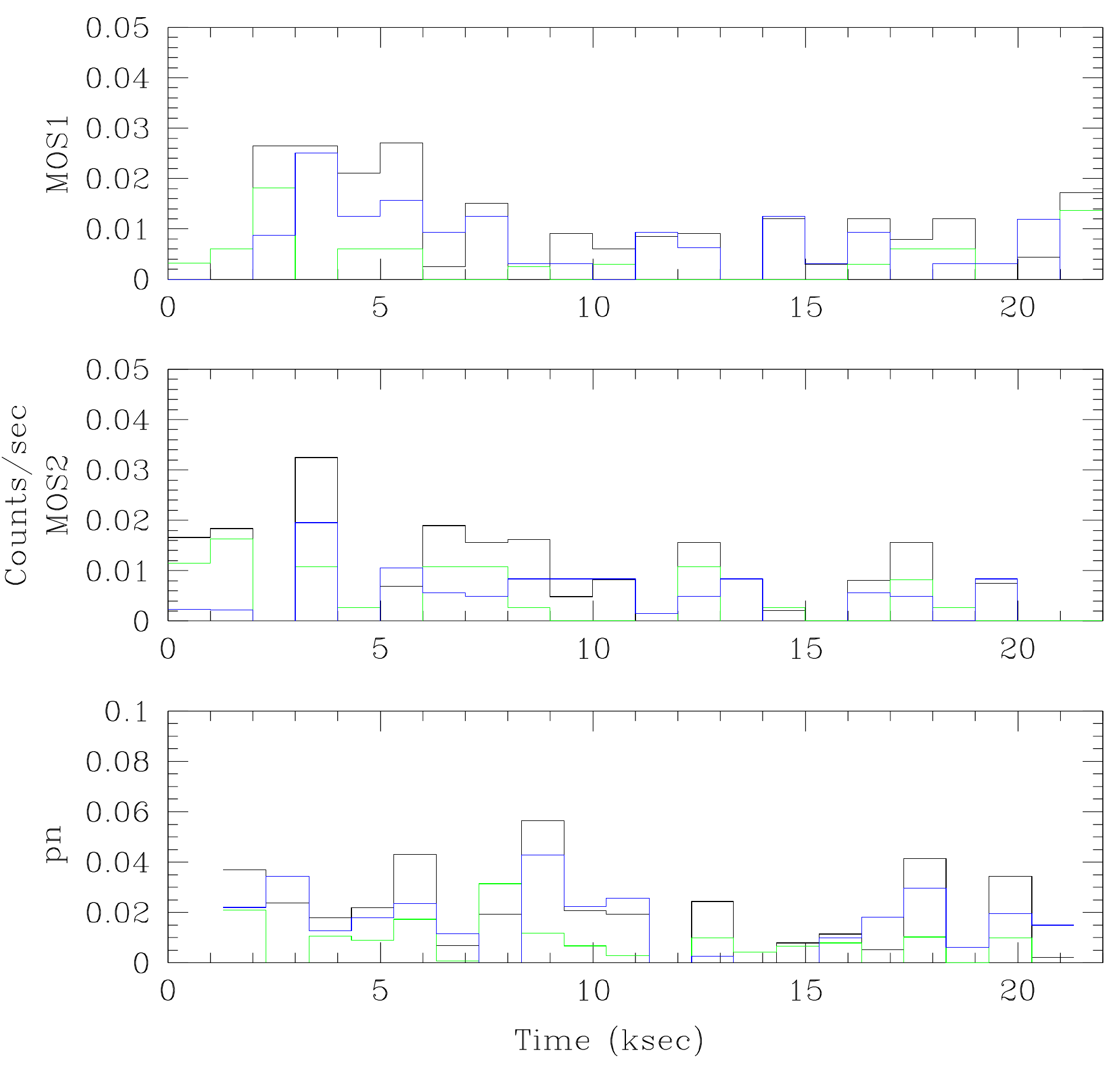}}
\end{center}
\end{minipage}
\begin{minipage}{5.6cm}
\begin{center}
\resizebox{5.6cm}{!}{\includegraphics{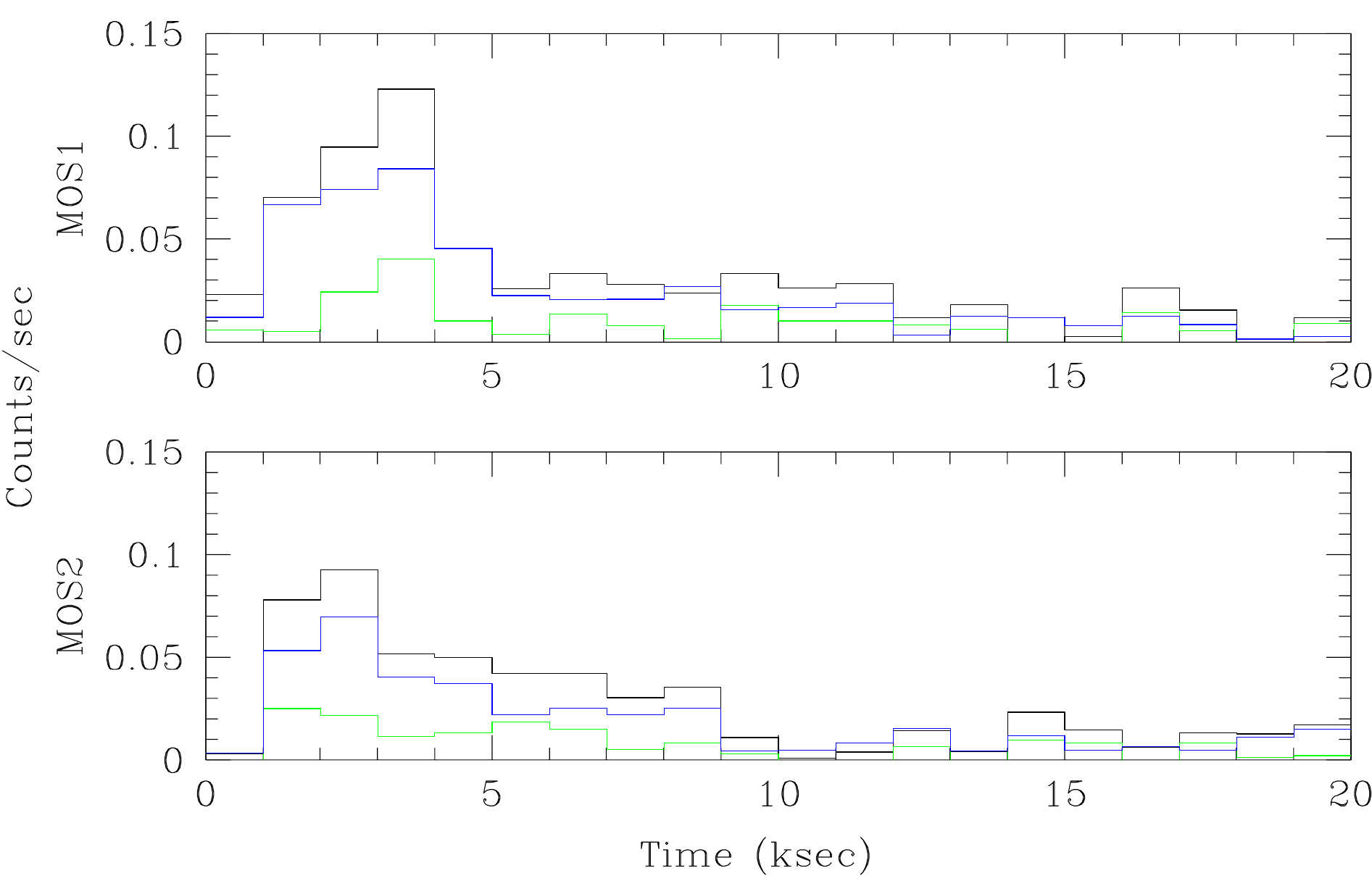}}
\end{center}
\end{minipage}
\hfill
\begin{minipage}{5.6cm}
\begin{center}
\resizebox{5.6cm}{!}{\includegraphics{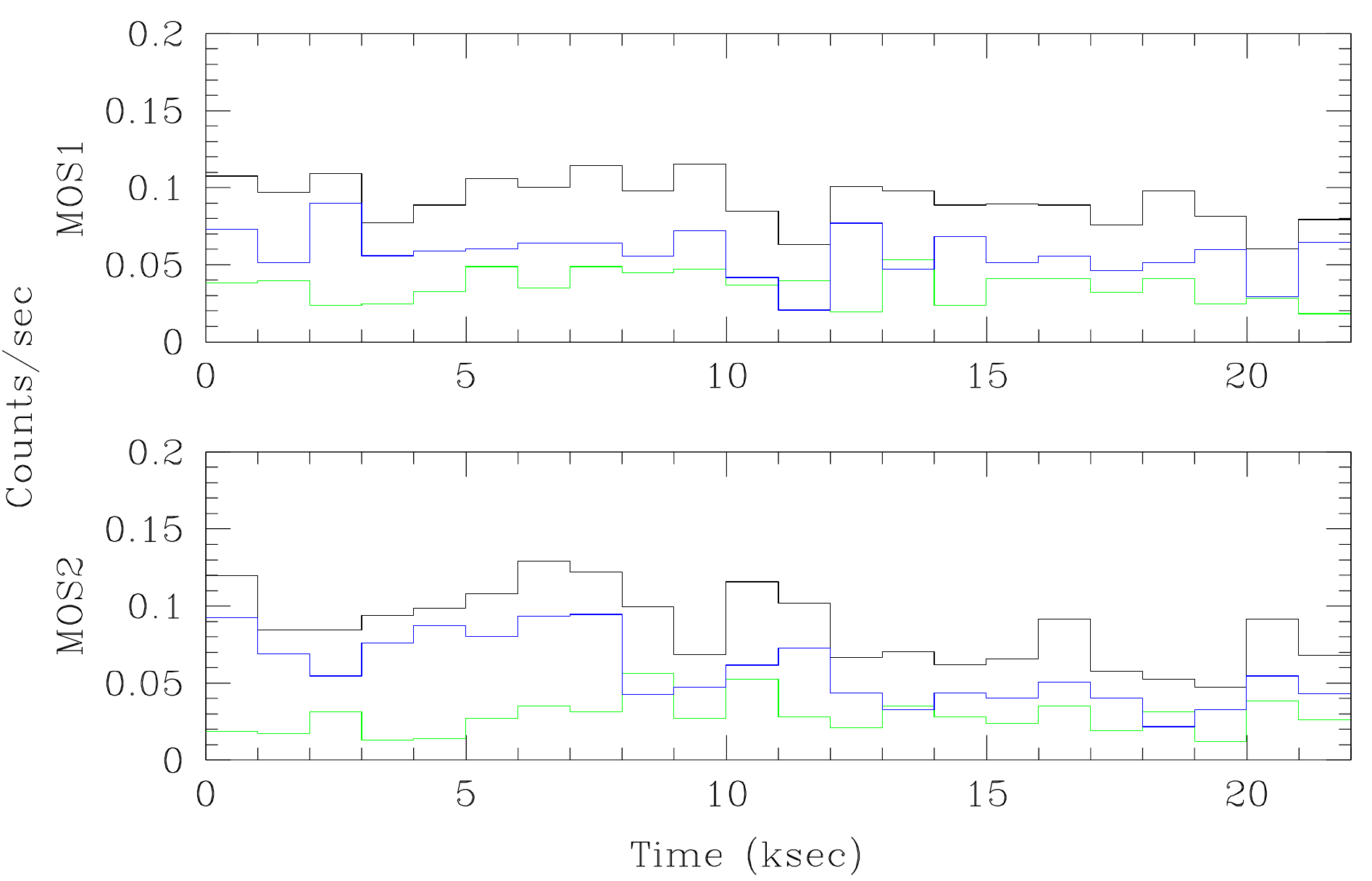}}
\end{center}
\end{minipage}
\hfill
\begin{minipage}{5.6cm}
\begin{center}
\end{center}
\end{minipage}
\caption{EPIC light curves of variable sources illustrating the variety of behaviours that are observed. The green, blue, and black histograms yield the background-subtracted light curves in the medium, hard, and total energy band respectively. The time is given in ksec from the start of the corresponding observation. Top, from left to right: sources 39, 132, and 190 during observations 4, 6, and 2, respectively. Bottom, from left to right: sources 67 and 176 during observations 1 and 2, respectively. \label{lcs}}
\end{figure*}

The variability comes in different flavours. In some cases, it is clearly associated with a large (a factor 10 or more) increase in the count rate during a specific observation (sources number 34, 39, 67, 132, and 173). In other cases, the count rates display a modulation of typically a factor 2-3 in certain energy bands (sources number 59, 107, 126, 144, 160, 182) from one observation to the other. In yet another group of (bright) objects, the variability is even more subtle, but still significant (CPR2002\,A11). We note that the list of variable sources, especially those of the last category, is not restricted to faint (probably low-mass) sources, but also includes OB-type stars. Table\,\ref{OBtable} indicates that 9 OB stars out of 25 were detected to be variable, at least in one energy band, by at least two of the three EPIC instruments. This is in line with the conclusion of Naz\'e (\cite{Naze}), who found that between one-third and two-thirds of the massive stars in the 2XMM catalogue display long-term variations. In our sample of massive stars, the variable sources include known colliding-wind binary systems (Cyg\,OB2 \#5, \#8a, \#9), radial velocity variables (MT91\,516, Kiminki et al.\ \cite{Kiminki}), as well as stars for which multiplicity remains currently undetermined (Cyg\,OB2 \#12, CPR2002\,A11, Cyg\,OB2 \#22, MT91\,455, MT91\,534). Among the eight X-ray brightest O-stars (those with mean pn count rates above 0.05\,cts\,s$^{-1}$), the only object that does not display significant variations in our data is Cyg\,OB2 \#7. This star in addition does not display significant radial velocity variations (Kiminki et al.\ \cite{Kiminki}). These results suggest that in our sample the X-ray variability of O-type stars could be an indicator of multiplicity.

Adopting a timebin of 1\,ksec, we extracted the light curves of those sources that display significantly higher count rates during a specific observation for each instrument and over three different energy bands (medium, hard, and total). A priori, these long-term variations could represent a flaring event during the observation with the highest count rate. X-ray flaring is a common property of magnetically active stars (e.g.\ Favata et al.\ \cite{Favata}). Albacete Colombo et al.\ (\cite{AC2}) detected flaring activity in about 15\% of the sources in their 100\,ksec {\it Chandra} observation of Cyg\,OB2. 

The X-ray light curves of the flare-candidates actually display a wide variety of behaviours (see Fig.\,\ref{lcs}): 
\begin{itemize}
\item[$\bullet$] Eight sources (34, 39, 64, 67, 86, 132, 149, and 173) display a genuine flare, although in two cases (64 and 86), our observations only cover the decay phase of an event that most likely started before our observation. We attempted to fit the decay phase of the light curves in the total energy band with an exponential decrease of the form $CR_{\rm max}\,\exp{\left[-\frac{t}{\tau}\right]}$, where $CR_{\rm max}$ is the count rate at the maximum of the flare and $\tau$ is the exponential decay time. The resulting decay times of five objects are between 1.7 and 7.3\,ksec (i.e.\ 0.5 and 2 hours). Albacete-Colombo et al.\ (\cite{AC2}) found decay times ranging from 0.5 to 10 hours in their 100\,ksec observation. Our dataset is clearly biased against the detection of flare events with decay times longer than about 25\,ksec (i.e.\ the mean duration of our individual observations). We note that some flare events cannot be modelled by an impulsive rise followed by an exponential decay. Source 173 for instance rises rather slowly (within 10\,ksec) to its maximum and remains there for the remainder of observation 5. The flare event of source 34 is probably an example of sustained heating with a second flare occuring about 6\,ksec after the first, more energetic, one (see Fig.\,\ref{spec10}).
\item[$\bullet$] Twenty-five sources (27, 57, 68, 70, 78, 82, 84, 89, 93, 94, 101, 106, 109, 124, 128, 130, 135, 137, 139, 140, 153, 179, 186, 190, and 196) display a light curve with no clear flare events, but with some short-term (a few ksec) variations. In most cases, a significant fraction (typically more than half) of the photons of these sources are emitted in the hard energy band. Two of these sources (70 and 78) have a higher count rate during two observations (3 + 5 and 2 + 5, respectively).  
\item[$\bullet$] Eight sources (16, 73, 79, 92, 134, 164, 169, and 176) display a high count rate during a specific observation with slow variations (i.e.\ on the timescale of the duration of the observation). These objects could be candidates for long duration (significantly longer than the duration of the observation) flares. 
\end{itemize}
 
\subsection{Spectra of variable non-OB sources}
\begin{figure*}[h!tb]
\begin{minipage}{8cm}
\begin{center}
\resizebox{8cm}{!}{\includegraphics{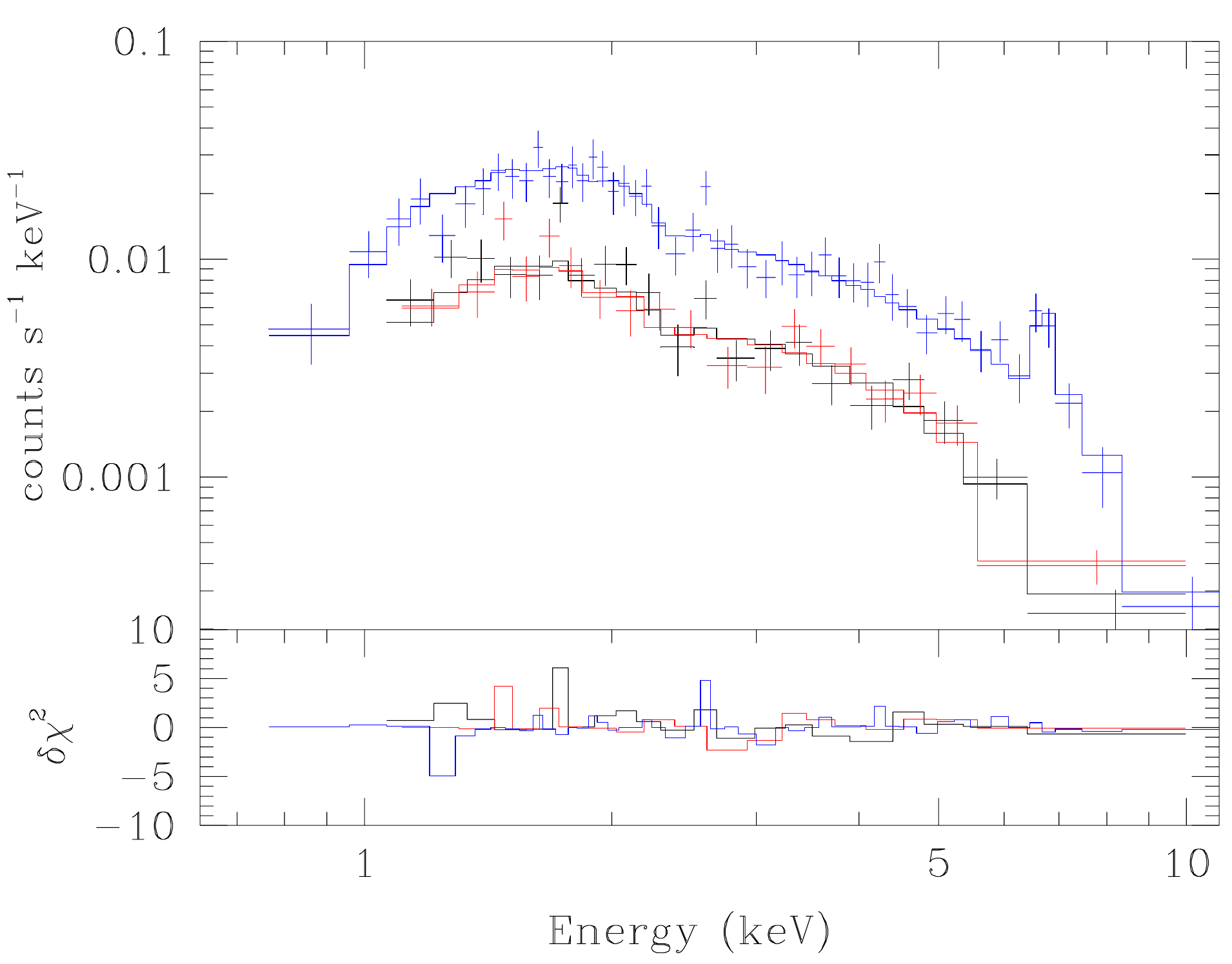}}
\end{center}
\end{minipage}
\hfill
\begin{minipage}{8cm}
\begin{center}
\resizebox{8cm}{!}{\includegraphics{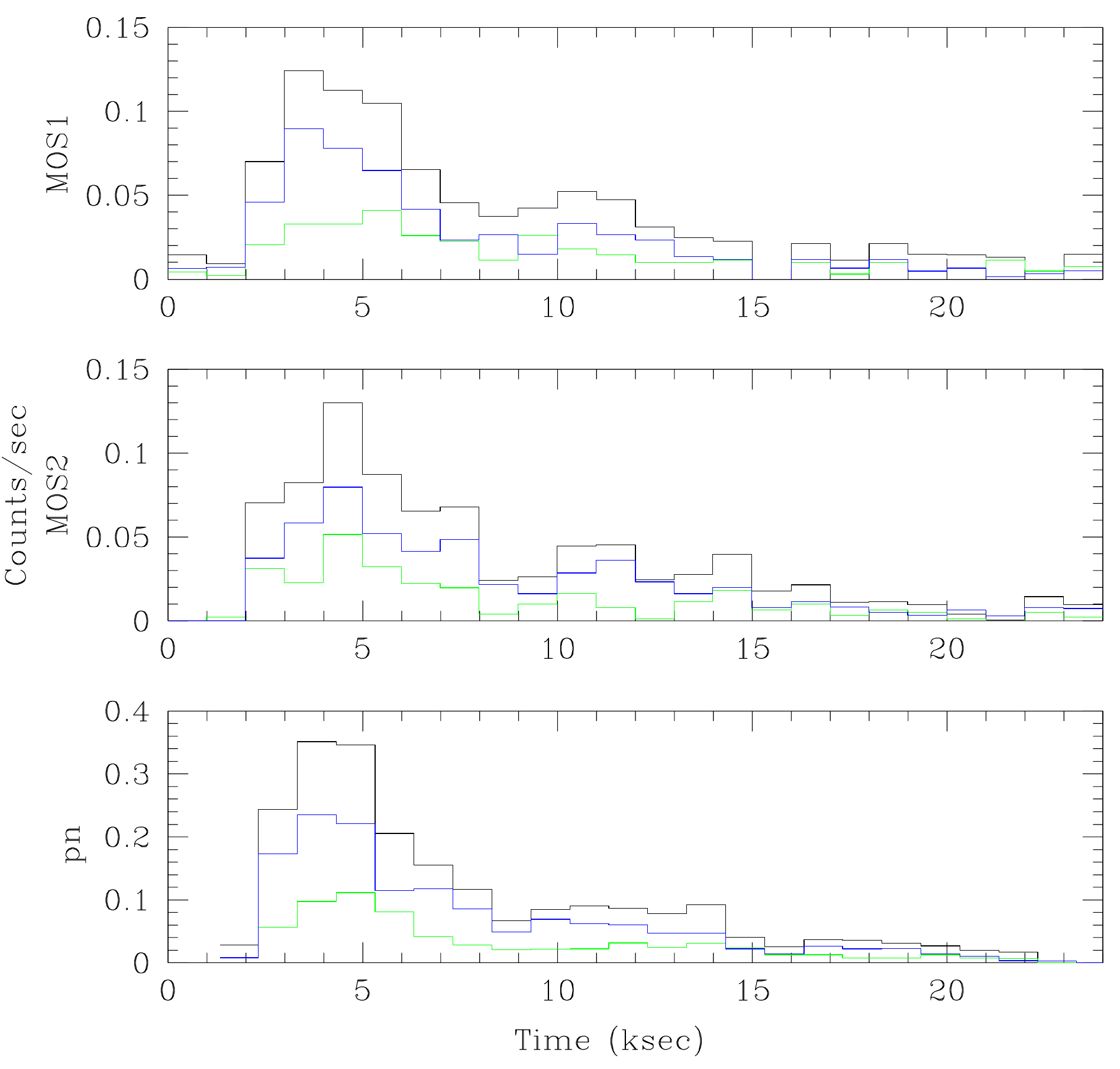}}
\end{center}
\end{minipage}
\caption{Left: EPIC spectra (along with the best-fit model) of source 34 during the third observation. Right: EPIC light curves of this source during the same observation. The green, blue, and black histograms indicate the light curves in the medium, hard, and total energy bands, respectively. The time is given in ksec elapsed from the start of the third observation.\label{spec10}}
\end{figure*}

As pointed out in Sect.\,\ref{lc}, a number of sources appear much brighter during a single observation than in average, although not all of these sources actually display a genuine flaring behaviour. For the brightest of these objects, we were able to extract and analyse their EPIC spectra. We attempted to fit these spectra with first an absorbed thermal plasma model and then an absorbed power-law model using the {\tt xspec} software (version 12.6.0, Arnaud \cite{Arnaud}). Except for a few cases (sources 73, 176)  where both models provide equally good fits, the absorbed thermal plasma model (wabs $\times$ apec) is the one that most closely fits the data. The results are summarized in Table\,\ref{specflare} and the fit of the spectra of source 34 is illustrated in Fig.\,\ref{spec10}.
\begin{table*}[htb]
\caption{Results of the spectral fits of the flaring sources with single temperature thermal plasma models. \label{specflare}}
\begin{center}
\begin{tabular}{c c c c c c c c c c}
\hline
\multicolumn{10}{c}{wabs $\times$ apec(T)}\\
\hline
Object & Obs. & Inst. & d.o.f. & $\chi^2$ & $N_H$                & $kT$  & norm & $f_X^{\rm obs}$ & $f_X^{\rm un}$ \\
       &      &       &        &          & ($10^{22}$\,cm$^{-2}$) & (keV) & ($10^{-14}$\,cm$^{-5}$) & (erg\,cm$^{-2}$\,s$^{-1}$) & (erg\,cm$^{-2}$\,s$^{-1}$) \\
\hline
 67 & 1 & M1+M2 & 21 & 0.92 & $1.22^{+.75}_{.52}$ & $12.2^{}_{-7.7}$ & $(3.6^{+1.5}_{-0.4}) \times 10^{-4}$ & $(5.24^{+0.57}_{-5.08}) \times 10^{-13}$ & $7.32 \times 10^{-13}$\\
 73 & 2  & EPIC & 11 & 1.27 & $0.90^{+.57}_{-.48}$ & $4.2^{+6.3}_{-1.6}$ & $(1.0^{+.4}_{-.3}) \times 10^{-4}$ & $(1.06^{+0.10}_{-0.39}) \times 10^{-13}$ & $1.64 \times 10^{-13}$\\
176 & 2 & M1+M2 & 38 & 1.22 & $1.10^{+.29}_{-.35}$ & $6.2^{+12.9}_{-2.1}$ & $(11.0^{+2.3}_{-1.9}) \times 10^{-4}$ & $(13.83^{+0.84}_{-1.69}) \times 10^{-13}$ & $20.46 \times 10^{-13}$\\
 34 & 3 & EPIC & 87 & 0.78 & $1.02^{+.12}_{-.12}$ & $8.1^{+2.6}_{-1.8}$ & $(4.4^{+.4}_{-.3}) \times 10^{-4}$ & $(6.17^{+0.22}_{-0.36}) \times 10^{-13}$ & $8.68 \times 10^{-13}$\\
 53 & 3 & pn & 30 & 0.75 & $0.26^{+.07}_{-.06}$ & $5.5^{+2.3}_{-.9}$ & $(1.9^{+.2}_{-.2}) \times 10^{-4}$ & $(2.86^{+0.24}_{-0.41}) \times 10^{-13}$ & $3.43 \times 10^{-13}$\\
 39 & 4 & EPIC & 34 & 1.15 & $1.17^{+.40}_{-.25}$ & $9.6^{+10.4}_{-4.6}$ & $(3.1^{+.7}_{-.3}) \times 10^{-4}$ & $(4.37^{+0.26}_{-0.71}) \times 10^{-13}$ & $6.16 \times 10^{-13}$\\
173 & 5 & EPIC & 29 & 1.49 & $1.13^{+.39}_{-.36}$ & $6.9^{+17.8}_{-2.7}$ & $(6.1^{+1.6}_{-1.0}) \times 10^{-4}$ & $(7.97^{+0.36}_{-1.73}) \times 10^{-13}$ & $11.62 \times 10^{-13}$\\
132 & 6 & M1 + M2 & 25 & 0.76 & $1.08^{+.34}_{-.29}$ & $6.9^{+11.4}_{-2.7}$ & $(4.5^{+1.0}_{-0.7}) \times 10^{-4}$ & $(5.94^{+0.40}_{-1.20}) \times 10^{-13}$ & $8.62 \times 10^{-13}$\\
\vspace*{-3mm}\\
\hline
\end{tabular}
\tablefoot{The last two columns yield the observed and absorption corrected flux over the 0.5 -- 10.0\,keV energy range.}
\end{center}
\end{table*}

All these sources have rather hard spectra characterized by high plasma temperatures ($kT > 4$\,keV) as expected for magnetically triggered X-ray flares. The column densities are somewhat lower than expected for Cyg\,OB2 members, especially in the case of source 53\footnote{We do not have a 2MASS counterpart with reliable photometry for this source, but both the OM photometry and the optical photometry from MT91 support the idea that this star is less heavily reddened than the majority of the stars towards Cyg\,OB2.}. However, we stress that, given the high plasma temperatures, the column densities are rather poorly constrained. We emphasize that the quoted fluxes correspond to average values over the entire observation under consideration. For flaring sources, they do not represent the peak luminosity. We note that the errors in the fluxes should be regarded only as indicative since the relevant {\tt xspec} routine does not account for possible correlations between model parameters.

\section{X-ray spectra of OB-type stars\label{specOB}}
At the sensitivity level of the current generation of X-ray telescopes, most early-type stars do not display short-term X-ray variability. However, in massive binary systems and single massive stars with magnetic fields of sufficient strength to confine the wind in the equatorial region, a modulation of the X-ray emission is expected and has been observed on the orbital or rotational period typically of the order of either days or weeks (see G\"udel \& Naz\'e \cite{GN} and references therein).

The X-ray spectra of three of the brightest sources, (Cyg\,OB2 \#5, \#8a, and \#9) were discussed elsewhere (De Becker et al.\ \cite{cyg8a1}, Linder et al.\ \cite{cyg5}, Naz\'e et al.\ \cite{cyg9}, and Blomme et al.\ \cite{cyg8a2}), and we do not repeat their analysis here. For the other OB-stars that are sufficiently X-ray bright, we analysed the spectra of each observation using {\tt xspec}. 

The interstellar neutral-hydrogen column density is evaluated from the $E(B - V)$ colour excess computed from the $B - V$ colour measured by Massey \& Thompson (\cite{massey}) and by adopting the gas-to-dust ratio of Bohlin et al.\ (\cite{Bohlin}). Early-type stars host dense stellar winds that can contribute to the absorption of X-rays produced in the inner regions of these winds. Therefore, whenever this turned out to be supported by our observations, we included an ionized wind absorption model (Naz\'e et al.\ \cite{wind}) in addition to the neutral interstellar-medium absorption.   

We compare the properties of these OB stars to the general X-ray emission properties derived by Naz\'e (\cite{Naze}), who analysed the data of OB stars in the 2XMM catalogue. Naz\'e (\cite{Naze}) obtained $\log{\frac{L_{\rm X}}{L_{\rm bol}}} = -6.45 \pm 0.51$ for those O-stars that have data of sufficient quality for a spectral fit and $\log{\frac{L_{\rm X}}{L_{\rm bol}}} = -6.97 \pm 0.40$ from 26 O-type stars in the region of Cyg\,OB2\footnote{Note that the latter relation was obtained by combining the data for stars for which the X-ray fluxes were either extracted from spectral fits or obtained from the EPIC count rates using an average conversion factor. Note also that these $\frac{L_{\rm X}}{L_{\rm bol}}$ relations include both single and binary O-stars as Naz\'e (\cite{Naze}) found no clear evidence of a general X-ray overluminosity of O + O binary systems. Some specific interacting-wind O-star binaries are indeed X-ray overluminous, but this overluminosity is not a general feature and is often restricted to specific orbital phases.}. In cases where a two-temperature model was required to fit the 2XMM EPIC spectra, the normalization of the hotter component was generally found to be less than 0.3 times that of the lower temperature component (Naz\'e \cite{Naze}). 
\subsection{Cyg\,OB2 \#12}
Massey \& Thompson (\cite{massey}) quote a colour of $B - V = 3.35$ for Cyg\,OB2 \#12, which corresponds to a colour excess of $E(B - V) \simeq 3.45$, assuming a B5\,I spectral type (see below). When we adopt a distance modulus of 10.4\,mag and $R_V = 3.1$, we find that $M_V = -9.64$. Assuming a bolometric correction of $-0.95$ then leads to $\log{\frac{L_{\rm bol}}{L_{\odot}}} = 6.13$. This luminosity places the star very near the empirical Humphreys-Davidson limit (Humphreys \& Davidson \cite{HD}).

The reddening of Cyg\,OB2 \#12 is larger than the average value of other OB stars in Cyg\,OB2, and the membership of this star is still uncertain. ESA's forthcoming {\it GAIA} mission should provide accurate astrometric measurements for this star that will help establish its membership. Spectral classifications of Cyg\,OB2 \#12 range from B3\,Ia to B8\,Ia, suggesting that the star actually changes its spectral type with time (Kiminki et al.\ \cite{Kiminki} and references therein). For instance, Kiminki et al.\ (\cite{Kiminki}) presented spectra that indicate a B3\,Iae type in September 2000 and B6-8 one year later. The nature of this star remains controversial, although based on its likely spectroscopic variability and incredibly high luminosity, Cyg\,OB2 \#12 has been suggested to be an LBV candidate (see Clark et al.\ \cite{CLA} and references therein). Cyg\,OB2 \#12 was detected as an {\it IRAS} source and is apparently surrounded by both warm and cold dust (Parthasarathy et al.\ \cite{Parta}). 

Souza \& Lutz (\cite{SL}) classified Cyg\,OB2 \#12 as a B8 supergiant with weak Balmer lines and emission in H$\alpha$ suggesting that the other Balmer absorptions could be filled-in by emission. They also reported a variable absorption bluewards of H$\alpha$ that they tentatively interpreted as a blueshifted H$\alpha$ absorption associated with an expanding shell at a velocity of 1400\,km\,s$^{-1}$. Klochkova \& Chentsov (\cite{KC}) classified Cyg\,OB2 \#12 as B5\,Ia$^+$ with an uncertainty of half a subtype. These authors presented evidence of a line radial-velocity gradient that might indicate an infall of matter. They estimated that the star has a radius of order 338\,$R_{\odot}$ and a mass-loss rate of $4 \times 10^{-5}$\,$M_{\odot}$\,yr$^{-1}$. Klochkova \& Chentsov (\cite{KC}) inferred a rather slow wind velocity of 150\,km\,s$^{-1}$, which would have important consequences for the understanding of the observed X-ray emission (see below). Such a slow wind would indeed be able to produce hard X-ray emission neither by means of wind embedded shocks nor head-on collisions of the wind from two hemispheres in a magnetically confined wind model. If Cyg\,OB2 \#12 were a binary system, then the wind of the secondary star, colliding with that of the B supergiant, would have to be much faster. We note however that Leitherer et al.\ (\cite{LHSW}) derived a significantly larger wind velocity of 1400\,km\,s$^{-1}$ from the H$\alpha$ line profile and that Wang \& Zhu (\cite{WZ}) reported an expanding shell at a velocity of 3100\,km\,s$^{-1}$. 

\begin{table*}[h!tb]
\caption{Best fits of the EPIC spectra of Cyg\,OB2 \#12 for photon energies above 0.4\,keV.\label{fit12}}
\begin{center}
\begin{tabular}{c c c c c c c c c c c}
\hline
\multicolumn{11}{c}{Cyg\,OB2 \#12, wabs $\times$ (apec(T$_1$) + apec(T$_2$))}\\
\hline
Obs. & Inst. & d.o.f. & $\chi^2$ & N$_{\rm H}$           & $kT_1$ & norm$_1$              & $kT_2$ & norm$_2$              & $f_X^{\rm obs}$            & $f_X^{\rm un}$ \\
     &       &        &          & ($10^{22}$\,cm$^{-2}$)& (keV)  & ($10^{-14}$\,cm$^{-5}$) & (keV)  & ($10^{-14}$\,cm$^{-5}$) & (erg\,cm$^{-2}$\,s$^{-1}$) & (erg\,cm$^{-2}$\,s$^{-1}$) \\
\hline
1 & M1+2 & 129 & 1.29 & $1.88^{+.06}_{-.06}$ & $0.74^{+.05}_{-.16}$ & $(7.36^{+1.14}_{-1.09}) \times 10^{-3}$ & $1.87^{+.18}_{-.13}$ & $(4.18^{+0.64}_{-1.09}) \times 10^{-3}$ & $(2.86^{+.04}_{-.09}) \times 10^{-12}$ & $2.36 \times 10^{-11}$ \\
2 & M1+2 & 148 & 1.34 & $1.83^{+.05}_{-.05}$ & $0.80^{+.05}_{-.06}$ & $(7.95^{+0.94}_{-0.52}) \times 10^{-3}$ & $2.19^{+.22}_{-.19}$ & $(3.76^{+0.64}_{-0.50}) \times 10^{-3}$ & $(3.28^{+.06}_{-.08}) \times 10^{-12}$ & $2.40 \times 10^{-11}$ \\
3 & EPIC & 255 & 1.72 & $1.85^{+.03}_{-.03}$ & $0.77^{+.03}_{-.04}$ & $(8.54^{+0.73}_{-0.71}) \times 10^{-3}$ & $1.91^{+.11}_{-.09}$ & $(4.34^{+0.42}_{-0.45}) \times 10^{-3}$ & $(3.27^{+.05}_{-.04}) \times 10^{-12}$ & $2.64 \times 10^{-11}$ \\
4 & EPIC & 192 & 1.26 & $1.83^{+.04}_{-.05}$ & $0.74^{+.04}_{-.02}$ & $(8.84^{+0.91}_{-0.88}) \times 10^{-3}$ & $1.81^{+.16}_{-.13}$ & $(3.36^{+0.54}_{-0.55}) \times 10^{-3}$ & $(2.74^{+.08}_{-.09}) \times 10^{-12}$ & $2.61 \times 10^{-11}$ \\
5 & EPIC & 108 & 1.15 & $2.06^{+.09}_{-.09}$ & $0.78^{+.10}_{-.08}$ & $(7.32^{+1.33}_{-1.25}) \times 10^{-3}$ & $2.28^{+.98}_{-.42}$ & $(1.85^{+0.77}_{-0.71}) \times 10^{-3}$ & $(1.99^{+.04}_{-.23}) \times 10^{-12}$ & $2.00 \times 10^{-11}$ \\
6 & EPIC & 140 & 1.10 & $2.05^{+.07}_{-.08}$ & $0.72^{+.08}_{-.14}$ & $(7.57^{+1.28}_{-1.16}) \times 10^{-3}$ & $2.10^{+.51}_{-.31}$ & $(2.62^{+0.93}_{-0.76}) \times 10^{-3}$ & $(2.17^{+.08}_{-.15}) \times 10^{-12}$ & $2.21 \times 10^{-11}$ \\
\vspace*{-3mm}\\
\hline
\end{tabular}
\end{center}
\tablefoot{The first three columns yield the number of the pointing, the instrument combination used in the fit and the number of degrees of freedom. The reduced $\chi^2$ is given in column 4, whilst the columns 5 to 9 provide the parameters of the absorption component and the emitting plasma. The last two columns yield the observed and absorption corrected flux over the 0.5 -- 10.0\,keV energy range.}
\end{table*}
\begin{figure*}[thb]
\begin{minipage}{8cm}
\begin{center}
\resizebox{8cm}{!}{\includegraphics{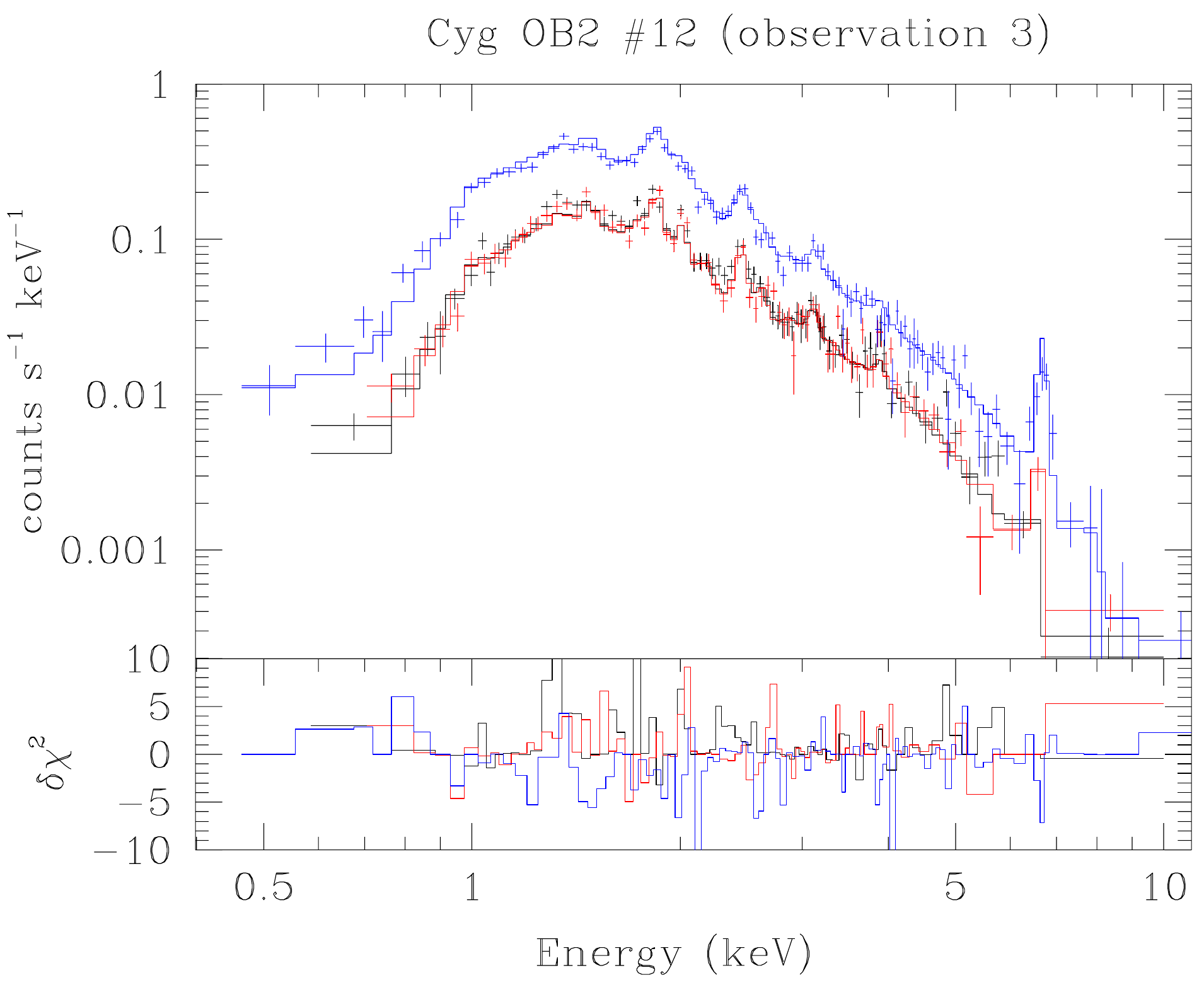}}
\end{center}
\end{minipage}
\hfill
\begin{minipage}{8cm}
\begin{center}
\resizebox{8cm}{!}{\includegraphics{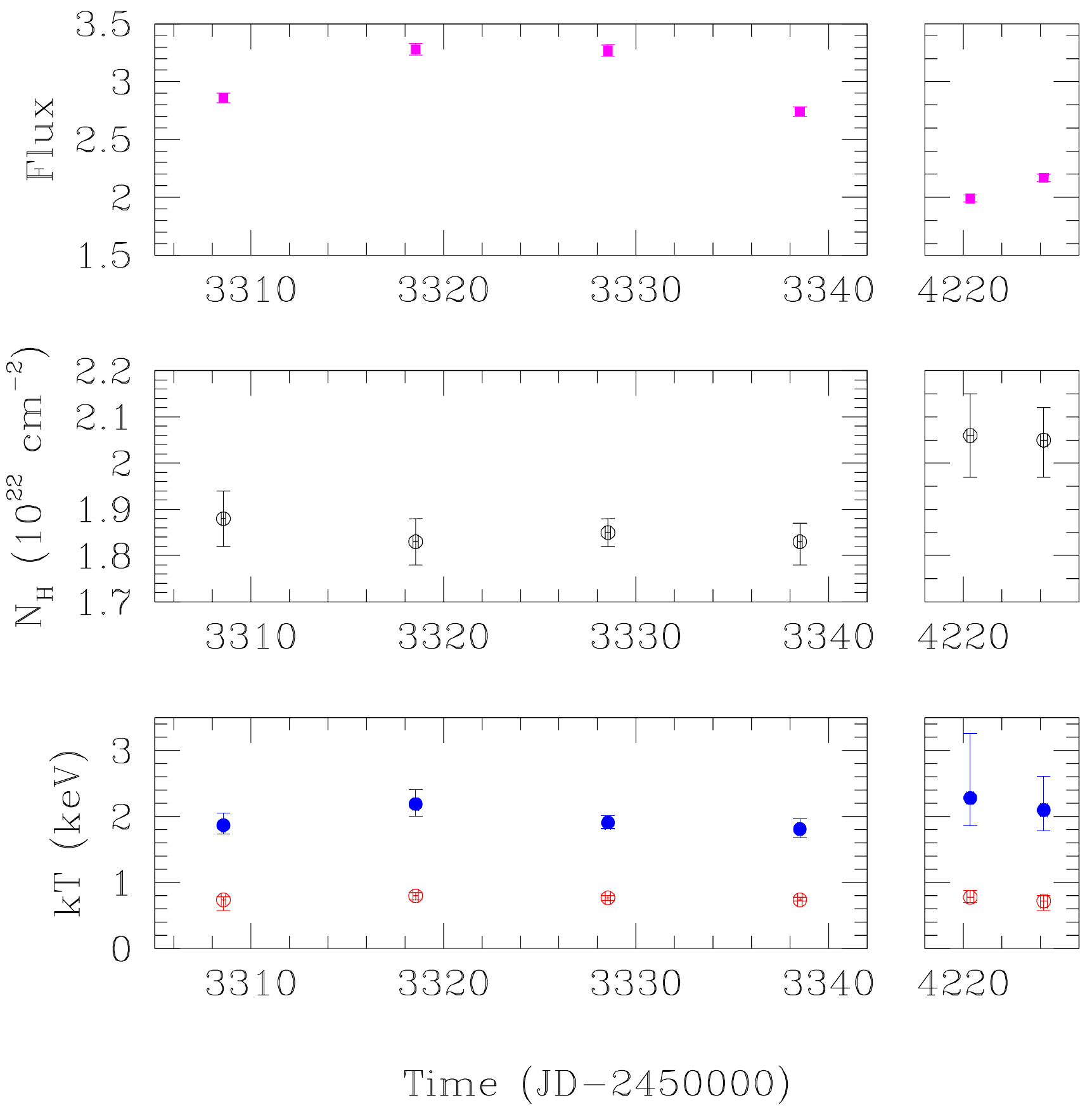}}
\end{center}
\end{minipage}
\caption{Left: EPIC spectra (along with the best-fit model) of Cyg\,OB2 \#12 as observed during the third observation. Right: Best-fit parameters of the EPIC spectra of Cyg\,OB2 \#12. The top panel illustrates the observed flux (in units $10^{-12}$\,erg\,cm$^{-2}$\,s$^{-1}$) over the 0.5 -- 10\,keV energy range. The middle panel yields the column density towards the X-ray emitting plasma, whilst the bottom panel displays the two temperatures of the emission components.\label{param12}}
\end{figure*}

The EPIC spectra of this star cannot be fitted with single-temperature plasma models and we instead used two-temperature models. The huge interstellar column density that we infer from the $E(B - V)$ colour excess (N$_{\rm H, ISM} = 2.0 \times 10^{22}$\,cm$^{-2}$) produces such a strong absorption at low photon energies that there is no room left for an additional ionized absorption column associated with the stellar wind. After testing several options of absorbed two plasma temperature models, we found that the optimal results are obtained when only interstellar absorption is included and the absorbing column is left as a free parameter. The results of the fits are quoted in Table\,\ref{fit12} and illustrated in Fig.\,\ref{param12}. The average absorption-corrected X-ray flux corresponds to an X-ray luminosity of $4.1 \times 10^{33}$\,erg\,s$^{-1}$. With the caveat that the large interstellar reddening renders the bolometric luminosity estimation and the X-ray flux correction more uncertain than for other stars in our sample, we derive $\log{\frac{L_{\rm X}}{L_{\rm bol}}} = -6.10$. This is an unusually large ratio, significantly larger than the average value for O-type stars and certainly exceptional for B-type stars. These characteristics imply that Cyg\,OB2 \#12 might either be a binary or host a stellar wind confined by a strong magnetic field. 

Figure\,\ref{param12} reveals moderate ($\sim 10$\%) flux variability on timescales from a few days to a few weeks, as well as large variations (37\%) on timescales of years\footnote{As a complement to our {\it XMM-Newton} data, we analysed the {\it ASCA}-SIS spectra of Cyg\,OB2 \#12 from 29 April 1993 previously analysed by Kitamoto \& Mukai (\cite{KM}). The SIS spectra can be fitted with a single temperature component at $kT \simeq 0.90$. The observed flux was found to be $2.2 \times 10^{-12}$\,erg\,cm$^{-2}$\,s$^{-1}$, quite similar to the value of our observation 6. However, owing to the modest angular resolution of {\it ASCA} and the contamination of the background by straylight from Cyg\,X-3, the uncertainty in this result is rather large (about 25\%). }. Long-term variations in the column density are also seen. However, there are apparently little or no changes in the plasma temperatures which are found to be $0.76 \pm 0.03$ and $2.03 \pm 0.19$\,keV for the cooler and hotter component respectively. Albacete-Colombo et al.\ (\cite{AC1}) reported a roughly linear decay of the ACIS-I countrate of Cyg\,OB2 \#12 from 0.18 to 0.16\,cts\,s$^{-1}$ (i.e.\ a 12\% variation) over their 100\,ksec {\it Chandra} observation. This result is very much in line with the $\sim 18$\% variations in the observed flux that we measure in our first four data points on timescales of about a week and the 10\% variations seen between the last two pointings. 

\begin{table}[htb]
\caption{OM $UVW1$ data of Cyg\,OB2 \#12. \label{S12}}
\begin{center}
\begin{tabular}{c c c c c}
\hline
JD-2450000 & $UVW1$ \\
\hline
$3308.505$ & $20.08 \pm 0.10$\\
$3318.478$ & $20.13 \pm 0.09$\\
$4220.282$ & $20.00 \pm 0.13$\\
$4224.061$ & $20.25 \pm 0.16$\\
\hline
\end{tabular}
\end{center}
\end{table}

Cyg\,OB2 \#12 was inside the OM FoV during four (1, 2, 5 and 6) out of the six observations. Owing to its heavy reddening, the star is detected only in the $UVW1$ band (see Table\,\ref{S12}) with an average magnitude of 20.11. The lack of significant variations in this magnitude with time, indicates that if changes in spectral type have occured during our campaign, they were not associated with large changes in the $UVW1$ spectral domain.  
\subsection{Cyg\,OB2 \#7}
This O3\,If$^*$ supergiant displays rather soft X-ray spectra that can be fitted well using a single temperature plasma component. From the colour excess $E(B - V) = 1.73$ (Massey \& Thompson \cite{massey}), we infer an interstellar column density of $N_{\rm H}^{\rm ISM} = 1.0 \times 10^{22}$\,cm$^{-2}$. The spectra show evidence of additional absorption and we thus included absorption by an ionized wind in our model. The results of the fit (at energies above 0.4\,keV) are listed in Table\,\ref{fit7}. There is no strong evidence of variability. The observed flux and the wind column density remain constant to within 10\%. At first sight, the plasma temperature varies between 0.56 and 0.77\,keV. However, there is some degeneracy in the fits between the plasma temperature and its normalization and all the spectra are actually consistent with each other.

\begin{table*}[htb]
\caption{Best-fits of the EPIC spectra of Cyg\,OB2 \#7 for photon energies above 0.4\,keV. \label{fit7}}
\begin{center}
\begin{tabular}{c c c c c c c c c}
\hline
\multicolumn{9}{c}{Cyg\,OB2 \#7, wabs $\times$ wind $\times$ apec(T)}\\
\hline
Obs. & Inst. & d.o.f. & $\chi^2$ & $\log{N^{\rm wind}}$ & $kT$ & norm                  & $f_X^{\rm obs}$ & $f_X^{\rm un}$ \\
     &       &        &          & (cm$^{-2}$)        & (keV) & ($10^{-14}$\,cm$^{-5}$) & (erg\,cm$^{-2}$\,s$^{-1}$) & (erg\,cm$^{-2}$\,s$^{-1}$) \\
\hline
1 & EPIC & 59 & 1.03 & $21.80^{+.06}_{-.06}$ & $0.61^{+.05}_{-.04}$ & $(1.27^{+0.24}_{-0.20}) \times 10^{-3}$ & $(1.73^{+.13}_{-.16}) \times 10^{-13}$ & $8.81 \times 10^{-13}$ \\
2 & M1+2 & 34 & 1.25 & $21.76^{+.08}_{-.08}$ & $0.76^{+.07}_{-.07}$ & $(0.89^{+0.17}_{-0.15}) \times 10^{-3}$ & $(1.78^{+.20}_{-0.21}) \times 10^{-13}$ & $7.53 \times 10^{-13}$ \\
3 & EPIC & 73 & 1.75 & $21.79^{+.05}_{-.06}$ & $0.60^{+.04}_{-.04}$ & $(1.30^{+0.21}_{-0.18}) \times 10^{-3}$ & $(1.70^{+.09}_{-.11}) \times 10^{-13}$ & $9.17 \times 10^{-13}$ \\
4 & M1+2 & 22 & 0.71 & $21.72^{+.17}_{-.16}$ & $0.56^{+.08}_{-.11}$ & $(1.39^{+1.15}_{-0.38}) \times 10^{-3}$ & $(1.75^{+.10}_{-.34}) \times 10^{-13}$ & $10.97 \times 10^{-13}$ \\
5 & M1+2 & 20 & 0.88 & $21.83^{+.10}_{-.11}$ & $0.77^{+.16}_{-.10}$ & $(1.00^{+0.32}_{-0.24}) \times 10^{-3}$ & $(1.96^{+.14}_{-0.48}) \times 10^{-13}$ & $7.60 \times 10^{-13}$ \\
6 & M1+2 & 28 & 1.14 & $21.84^{+.09}_{-.10}$ & $0.58^{+.11}_{-.08}$ & $(1.40^{+0.61}_{-0.38}) \times 10^{-3}$ & $(1.66^{+.13}_{-.32}) \times 10^{-13}$ & $8.49 \times 10^{-13}$ \\
\vspace*{-3mm}\\
\hline
\end{tabular}
\end{center}
\tablefoot{The fit was perfomed with $N_{\rm H}^{\rm ISM}$ fixed to $1.0 \times 10^{22}$\,cm$^{-2}$. The first three columns yield the number of the pointing, the instrument combination used in the fit, and the number of degrees of freedom. The reduced $\chi^2$ is given in column 4, whilst the columns 5 to 9 provide the parameters of the absorption component and the emitting plasma. The last two columns yield the observed and ISM absorption corrected fluxes over the 0.5 -- 10.0\,keV energy range.}
\end{table*}

Herrero et al.\ (\cite{HPN}) inferred a luminosity of $\log{\frac{L_{\rm bol}}{L_{\odot}}} = 5.91 \pm 0.10$ (assuming the distance modulus of Massey \& Thompson \cite{massey}). Scaling this number to the lower distance modulus of Hanson (\cite{hanson}, $DM = 10.4$) that we have adopted in this paper, yields $\log{\frac{L_{\rm bol}}{L_{\odot}}} = 5.59$. Proceeding in the same way with the bolometric magnitude derived by Negueruela et al.\ (\cite{Negu}) yields $\log{\frac{L_{\rm bol}}{L_{\odot}}} = 5.87$. Our mean, ISM absorption-corrected X-ray flux $(8.76 \pm 1.26)\,10^{-13}$\,erg\,cm$^{-2}$\,s$^{-1}$ yields $\log{\frac{L_{\rm X}}{L_{\rm bol}}} = -6.99 \pm 0.06$ or $-7.27 \pm 0.06$ for the Herrero et al.\ (\cite{HPN}) and Negueruela et al.\ (\cite{Negu}) bolometric luminosities, respectively. We conclude that the X-ray emission of Cyg\,OB2 \#7 does not reveal any behaviour (hard X-ray emission, strong variability, or X-ray overluminosity) that could hint at binarity or the effects of magnetic confinement of the stellar wind.

\subsection{MT91\,516}
This star is of spectral type O5.5\,V((f)). Its fundamental parameters were determined by Herrero et al.\ (\cite{HCVM}). Scaled to the distance adopted in the present paper, the bolometric luminosity is equal to $\log{\frac{L_{\rm bol}}{L_{\odot}}} = 6.01$. Proceeding in the same way with the bolometric magnitude inferred by Negueruela et al.\ (\cite{Negu}) yields $\log{\frac{L_{\rm bol}}{L_{\odot}}} = 6.12$.

From the colour excess $E(B - V) = 2.48$ (Massey \& Thompson \cite{massey}), we infer a large interstellar column density of $N_{\rm H}^{\rm ISM} = 1.44 \times 10^{22}$\,cm$^{-2}$. Nevertheless, the spectra display evidence of additional absorption that we modelled by including an ionized wind component. To achieve a reasonably good fit to the EPIC spectra, a two-temperature plasma model is required. The results of the fit to the EPIC spectra of MT91\,516 (at energies above 0.4\,keV) are listed in Table\,\ref{fit516}. 

There is a degeneracy between the value of the ionized-wind column density and the temperature of the softer plasma component, and our best-fit solution oscillates between $kT_1$ near 0.7\,keV with a very low wind column density (observations 2 and 4), and $kT_1$ near 0.25\,keV with a significantly larger wind column density (all other observations). The second temperature component at $kT_2 = (2.0 \pm 0.2)$\,keV is more tightly constrained. 

\begin{table*}[htb]
\caption{Same as Table\,\ref{fit7}, but for MT91\,516. \label{fit516}}
\begin{center}
\begin{tabular}{c c c c c c c c c c c}
\hline
\multicolumn{11}{c}{MT91\,516, wabs $\times$ wind $\times$ (apec(T$_1$) + apec(T$_2$))}\\
\hline
Obs. & Inst. & d.o.f. & $\chi^2$ & $\log{N^{\rm wind}}$ & $kT_1$ & norm$_1$              & $kT_2$ & norm$_2$              & $f_X^{\rm obs}$            & $f_X^{\rm un}$ \\
     &       &        &          & (cm$^{-2}$)        & (keV)  & ($10^{-14}$\,cm$^{-5}$) & (keV)  & ($10^{-14}$\,cm$^{-5}$) & (erg\,cm$^{-2}$\,s$^{-1}$) & (erg\,cm$^{-2}$\,s$^{-1}$) \\
\hline
1 & EPIC & 57 & 1.18 & $21.5^{+.3}_{-.8}$ & $0.28^{+.09}_{-.08}$ & $(5.72^{+18.14}_{-3.72}) \times 10^{-3}$ & $2.22^{+.37}_{-.31}$ & $(6.32^{+1.30}_{-0.90}) \times 10^{-4}$ & $(4.22^{+0.21}_{-0.74}) \times 10^{-13}$ & $4.55 \times 10^{-12}$ \\
2 & EPIC & 82 & 0.84 & $19.6^{+1.6}_{-.60}$ & $0.77^{+.22}_{-.15}$ & $(6.98^{+2.93}_{-0.76}) \times 10^{-4}$ & $2.26^{+.45}_{-.32}$ & $(6.00^{+0.64}_{-0.50}) \times 10^{-4}$ & $(4.97^{+0.03}_{-3.93}) \times 10^{-13}$ & $2.48 \times 10^{-12}$ \\
3 & EPIC & 52 & 1.2 & $21.7^{+.2}_{-.5}$ & $0.22^{+.10}_{-.06}$ & $(14.5^{+75.8}_{-11.2}) \times 10^{-3}$ & $1.91^{+.32}_{-.30}$ & $(8.85^{+2.04}_{-4.75}) \times 10^{-4}$ & $(4.58^{+0.25}_{-1.06}) \times 10^{-13}$ & $7.04 \times 10^{-12}$ \\
4 & EPIC & 53 & 1.09 & $19.0^{+2.6}_{-0.0}$ & $0.61^{+.16}_{-.20}$ & $(8.19^{+8.37}_{-2.57}) \times 10^{-4}$ & $1.86^{+.30}_{-.24}$ & $(7.25^{+4.53}_{-1.29}) \times 10^{-4}$ & $(4.73^{+0.01}_{-3.96}) \times 10^{-13}$ & $2.97 \times 10^{-12}$ \\
5 & EPIC & 20 & 1.13 & $22.1^{+.3}_{-.4}$ & $0.27^{+.35}_{-.10}$ & $(18.0^{+515.0}_{-17.4}) \times 10^{-3}$ & $1.85^{+1.09}_{-.51}$ & $(8.99^{+7.19}_{-4.36}) \times 10^{-4}$ & $(4.14^{+0.01}_{-3.24}) \times 10^{-13}$ & $2.08 \times 10^{-12}$ \\
6 & EPIC & 35 & 0.88 & $21.7^{+.2}_{-.6}$ & $0.32^{+.17}_{-.13}$ & $(4.84^{+33.91}_{-3.47}) \times 10^{-3}$ & $1.90^{+.66}_{-.42}$ & $(6.72^{+2.59}_{-1.86}) \times 10^{-4}$ & $(3.96^{+0.03}_{-1.36}) \times 10^{-13}$ & $3.14 \times 10^{-12}$ \\
\vspace*{-3mm}\\
\hline
\end{tabular}
\tablefoot{The fit was perfomed with $N_{\rm H}^{\rm ISM}$ fixed to  $1.44 \times 10^{22}$\,cm$^{-2}$.}
\end{center}
\end{table*}

The MOS1 and pn count rates of this source are found to be variable though with a moderate amplitude. The same kind of variability is found in the observed flux of the source, which displays variations with an amplitude of about 22\% (peak to peak). Owing to the degeneracy of $kT_1$, the ISM absorption-corrected flux displays very large variations. The mean value of the ISM absorption-corrected X-ray flux $(3.71 \pm 1.84)\,10^{-12}$\,erg\,cm$^{-2}$\,s$^{-1}$ yields $\log{\frac{L_{\rm X}}{L_{\rm bol}}} = -6.79 \pm 0.22$ (for the Herrero et al.\ \cite{HCVM} bolometric luminosity). Within the (large) errorbars, we find no significant X-ray overluminosity, relative to the canonical $\frac{L_{\rm X}}{L_{\rm bol}}$ value (Naz\'e \cite{Naze}). 

The star displays radial velocity variations with a peak-to-peak amplitude of 113\,km\,s$^{-1}$ (Kiminki et al.\ \cite{Kiminki}). Although these variations have an 11.5\% probability of being caused by random noise, they might also represent binarity. Mason et al.\ (\cite{Mason}) reported the detection of an astrometric companion of similar brightness (magnitude difference of 0.4) at a separation of 0.7\,arcsec. There is currently no evidence of any physical association between the two stars. Since there was no change in the position angle and separation of this pair over the time-frame of our X-ray observing campaign (Mason et al.\ \cite{Mason}), it is unlikely that the observed X-ray variability, and by virtue the X-ray emission itself, could be associated with interactions between the  astrometric components. 

\subsection{The pair Cyg\,OB2 \#22 A + B}
Cyg\,OB2 \#22 was classified as an O4\,III(f) giant and a fit to its spectrum yielded $\log{\frac{L_{\rm bol}}{L_{\odot}}} = 6.04$ (from the results of Herrero et al.\ \cite{HCVM}, scaled to our distance). The bolometric magnitude given by Negueruela et al.\ (\cite{Negu}) yields a slightly larger $\log{\frac{L_{\rm bol}}{L_{\odot}}} = 6.10$. However, Cyg\,OB2 \#22 was subsequently reported to be a visual binary system consisting of an O3\,I (star A) and an O6\,V star (star B, Walborn et al.\ \cite{Walborn1}, Mason et al.\ \cite{Mason}) separated by about 1.5\,arcsec. The O6\,V component is itself a double system with a separation of 0.2\,arcsec and a magnitude difference of 2.34 in the $z$ filter (Sota et al.\,\cite{Sota}). A priori, this result casts some doubt on the luminosity inferred by Herrero et al.\ (\cite{HCVM}), although the sum of the `typical' luminosities of an O3\,I and an O6\,V star (Martins et al.\ \cite{Martins}) amounts to a very similar value $\log{\frac{L_{\rm bol}}{L_{\odot}}} = 6.07$. In the following, we adopt this latter value. We note that the separation between the A and B components is too small to resolve their X-ray emission with {\it XMM-Newton}.

From the $B - V = 2.04$ colour index (Massey \& Thompson \cite{massey}), we derive an interstellar column of $N_{\rm H}^{\rm ISM} = 1.35 \times 10^{22}$\,cm$^{-2}$. To achieve a good fit of the EPIC spectra, we need to include an ionized wind absorption. In addition, a second plasma temperature is required for the first four observations. In the last two observations, which have lower fluxes and lower signal-to-noise ratio, this second temperature is not constrained and is actually not required to achieve an acceptable fit. 
\begin{table*}[htb]
\caption{Same as Table\,\ref{fit7}, but for Cyg\,OB2 \#22. \label{fit22}}
\begin{center}
\begin{tabular}{c c c c c c c c c c c}
\hline
\multicolumn{11}{c}{Cyg\,OB2 \#22, wabs $\times$ wind $\times$ (apec(T$_1$) + apec(T$_2$))}\\
\hline
Obs. & Inst. & d.o.f. & $\chi^2$ & $\log{N^{\rm wind}}$ & $kT_1$ & norm$_1$              & $kT_2$ & norm$_2$              & $f_X^{\rm obs}$            & $f_X^{\rm un}$ \\
     &       &        &          & (cm$^{-2}$)        & (keV)  & ($10^{-14}$\,cm$^{-5}$) & (keV)  & ($10^{-14}$\,cm$^{-5}$) & (erg\,cm$^{-2}$\,s$^{-1}$) & (erg\,cm$^{-2}$\,s$^{-1}$) \\
\hline
1 & EPIC & 62 & 1.18 & $21.6^{+.1}_{-.2}$ & $0.50^{+.09}_{-.13}$ & $(1.88^{+2.05}_{-.60}) \times 10^{-3}$ & $2.27^{+1.27}_{-.55}$ & $(2.10^{+1.28}_{-.88}) \times 10^{-4}$ & $(2.64^{+0.20}_{-0.73}) \times 10^{-13}$ & $1.99 \times 10^{-12}$ \\
2 & EPIC & 56 & 0.95 & $21.6^{+0.1}_{-.3}$ & $0.57^{+.09}_{-.10}$ & $(1.51^{+0.76}_{-0.42}) \times 10^{-3}$ & $3.33^{+2.5}_{-.88}$ & $(2.52^{+0.97}_{-0.92}) \times 10^{-4}$ & $(3.48^{+0.19}_{-0.86}) \times 10^{-13}$ & $1.94 \times 10^{-12}$ \\
3 & EPIC & 70 & 1.18 & $21.7^{+0.1}_{-.3}$ & $0.30^{+.10}_{-.11}$ & $(3.16^{+3.96}_{-1.88}) \times 10^{-3}$ & $0.89^{+.10}_{-.11}$ & $(6.96^{+2.68}_{-1.56}) \times 10^{-4}$ & $(1.99^{+0.12}_{-0.55}) \times 10^{-13}$ & $2.15 \times 10^{-12}$ \\
4 & EPIC & 37 & 1.69 & $21.8^{+0.2}_{-.3}$ & $0.29^{+.04}_{-.10}$ & $(6.51^{+6.97}_{-5.39}) \times 10^{-3}$ & $1.62^{+.85}_{-.38}$ & $(4.00^{+2.85}_{-1.95}) \times 10^{-4}$ & $(2.41^{+0.10}_{-1.13}) \times 10^{-13}$ & $2.62 \times 10^{-12}$ \\
5 & M1+M2 & 9 & 1.45 & $21.7^{+.3}_{-}$ & $0.93^{+.32}_{-.22}$ & $(0.91^{+.68}_{-.36}) \times 10^{-3}$ & -- & -- & $(1.96^{+0.18}_{-1.00}) \times 10^{-13}$ & $0.88 \times 10^{-12}$ \\
6 & EPIC & 22 & 1.23 & $21.5^{+.3}_{-.4}$ & $0.73^{+.32}_{-.14}$ & $(1.05^{+.84}_{-.48}) \times 10^{-3}$ & -- & -- & $(1.77^{+0.16}_{-0.60}) \times 10^{-13}$ & $1.31 \times 10^{-12}$ \\
\vspace*{-3mm}\\
\hline
\end{tabular}
\end{center}
\tablefoot{The fit was perfomed with $N_{\rm H}^{\rm ISM}$ fixed to $1.35 \times 10^{22}$\,cm$^{-2}$.}
\end{table*}

The observed flux varies by nearly a factor of two between the second and the sixth observations. This variability is not due to an instrumental effect as it is also seen in the count rates of all the individual EPIC cameras. The EPIC count rates of this source are indeed found to be variable at the 99.9\% level (though not in all energy bands for each instrument). The ISM absorption-corrected fluxes vary by a factor of three and the extreme values correspond to $L_{\rm X}$ in the range from $1.5 \times 10^{32}$ to $4.5 \times 10^{32}$\,erg\,s$^{-1}$ and $\log{\frac{L_{\rm X}}{L_{\rm bol}}}$ between $-7.47$ and $-7.00$. The star hence varies from underluminous by a factor three to just at the typical X-ray luminosity level. 

\subsection{CPR2002\,A11}
Negueruela et al.\ (\cite{Negu}) inferred $\log{\frac{L_{\rm bol}}{L_{\odot}}} = 5.44$ (again scaled to the distance adopted in the present paper) for this O7.5\,Ib-II(f) star. From $B - V = 2.19$ (for MT91\,267), we estimate that $N_{\rm H}^{\rm ISM} = 1.43 \times 10^{22}$\,cm$^{-2}$. The spectra are unlikely to contain an additional (wind) absorption component, thus we fitted them using a single-temperature plasma model with fixed interstellar absorption. 
\begin{table*}[htb]
\caption{Same as Table\,\ref{fit7}, but for CPR2002\,A11. \label{fitA11}}
\begin{center}
\begin{tabular}{c c c c c c c c c c c}
\hline
\multicolumn{8}{c}{CPR2002\,A11, wabs $\times$ apec(T)}\\
\hline
Obs. & Inst. & d.o.f. & $\chi^2$ & $kT$  & norm                  & $f_X^{\rm obs}$ & $f_X^{\rm un}$ \\
     &       &        &          & (keV) & ($10^{-14}$\,cm$^{-5}$) & (erg\,cm$^{-2}$\,s$^{-1}$) & (erg\,cm$^{-2}$\,s$^{-1}$) \\
\hline
1 & EPIC & 85 & 1.60 & $1.63^{+.07}_{-.06}$ & $(1.20^{+.05}_{-.04}) \times 10^{-3}$ & $(5.29^{+.23}_{-.21}) \times 10^{-13}$ & $1.70 \times 10^{-12}$ \\
2 & EPIC & 52 & 1.27 & $1.49^{+.09}_{-.09}$ & $(0.60^{+.03}_{-.04}) \times 10^{-3}$ & $(2.44^{+.16}_{-.17}) \times 10^{-13}$ & $0.88 \times 10^{-12}$ \\
3 & M1+M2 & 51 & 1.89 & $1.52^{+.09}_{-.08}$ & $(1.19^{+.07}_{-.08}) \times 10^{-3}$ & $(4.96^{+.26}_{-.28}) \times 10^{-13}$ & $1.74 \times 10^{-12}$ \\
4 & M1+M2 & 25 & 1.67 & $1.89^{+.19}_{-.20}$ & $(1.05^{+.08}_{-.08}) \times 10^{-3}$ & $(5.21^{+.43}_{-.44}) \times 10^{-13}$ & $1.43 \times 10^{-12}$ \\
5 & M1+M2 & 24 & 1.07 & $1.57^{+.15}_{-.14}$ & $(1.56^{+.11}_{-.12}) \times 10^{-3}$ & $(6.62^{+.71}_{-.50}) \times 10^{-13}$ & $2.22 \times 10^{-12}$ \\
6 & M1+M2 & 22 & 0.94 & $1.49^{+.15}_{-.17}$ & $(0.97^{+.08}_{-.08}) \times 10^{-3}$ & $(3.95^{+.33}_{-.34}) \times 10^{-13}$ & $1.42 \times 10^{-12}$ \\
\vspace*{-3mm}\\
\hline
\end{tabular}
\end{center}
\tablefoot{The fit was perfomed with $N_{\rm H}^{\rm ISM}$ fixed to $1.43 \times 10^{22}$\,cm$^{-2}$.}
\end{table*}
 
This source is found to be highly variable in count rate (in all three EPIC instruments) and this variability is reflected in the observed flux that varies by a factor of 2.7. The ISM absorption-corrected fluxes yield $\log{\frac{L_{\rm X}}{L_{\rm bol}}}$ between $-6.84$ and $-6.44$. This star hence appears somewhat overluminous in X-rays. Furthermore, that its X-ray emission varies and is dominated by a relatively hot plasma at $(1.60 \pm 0.15)$\,keV suggests that it might either be an interacting wind binary system or a star containing a magnetically confined wind. Spectroscopic monitoring in the optical is highly desirable to help us establish the multiplicity of this object.

\section{Conclusions\label{conclusion}}
We have discussed the X-ray properties of a number of stars in the core of Cyg\,OB2. One of this association's most interesting stars is the LBV candidate Cyg\,OB2 \#12. We have shown for the first time that the X-ray flux of this star varies on many different timescales ranging from days to years. These variations are mainly due to variations in the column density and changes in the emission measure of the harder component of the spectrum as one would expect for models of either an eccentric colliding wind binary or a rotationally modulated magnetically confined wind. Given that the variability occurs on timescales as short as days, the latter scenario seems slightly more plausible. An extensive monitoring of this star, both in the optical and the X-ray domain could shed new light on the nature of this intriguing object. 

Apart from the known colliding-wind systems, the majority of the O-type stars in Cyg\,OB2 appear to have rather normal X-ray properties. Their comparatively hard X-ray emission is not intrinsic to the sources, but is caused instead by the huge interstellar absorption towards the association. Several objects feature a second thermal component that is too hot to be easily explained by the wind-embedded shocks scenario. However, it must be stressed that this result is not restricted to the stars in our sample, nor direct evidence of either multiplicity or a magnetic field. Indeed, a second thermal component with a temperature around 2\,keV is frequently required to fit high quality CCD X-ray spectra of O-type stars, regardless of their multiplicity (e.g.\ Naz\'e et al.\ \cite{CCCP}). In our sample, the exceptions to this general statement are MT91\,516 and CPR2002\,A11, which we propose to be good candidates for binary systems. About one third of the OB stars display variability between our pointings. All known binary systems (except for the W UMa system Cyg\,OB2 \#27) are highly variable, which suggests that the other variable OB stars might also be binaries. However, we stress that these objects do not necessarily display a permanent excess X-ray emission; the Trapezium-like system Cyg\,OB2 \#22 even appears to be `X-ray underluminous' in some of our observations. 

Finally, a large number of the secondary sources, the bulk of which are most likely associated with pre-main sequence low-mass stars, also display highly significant variability. However, this variability appears to be more complex than the impulsive rise followed by an exponential decay that is commonly seen in flaring stars. Clarifying the relationship between these objects and the massive star population in Cyg\,OB2 will help us determine the impact of massive stars on star formation in their surroundings. In this respect, the {\it Chandra} very large program survey of Cyg\,OB2 (Drake \cite{Drake}) and its multi-wavelength support observations will certainly help us make substantial progress.   

\acknowledgement{My heartiest thanks go to Dr.\ Ya\"el Naz\'e for sharing her spectral extraction routines with me and for her help in preparing Fig.\,\ref{OMimage}. I thank the referee, Dr.\ Marc Gagn\'e, and the editor, Prof. Ralf Napiwotzki, for their reports that helped improve the manuscript and Dr.\ Nicolas Wright for stimulating discussion on Cyg\,OB2. I acknowledge support from the Fonds de Recherche Scientifique (FRS/FNRS), through the XMM/INTEGRAL PRODEX contract as well as by the Communaut\'e Fran\c caise de Belgique - Action de recherche concert\'ee - Acad\'emie Wallonie - Europe. This research made use of data products from the Two Micron All Sky Survey, which is a joint project of the University of Massachusetts and the Infrared Processing and Analysis Center/California Institute of Technology, funded by NASA and NSF.}

\end{document}